\documentclass[aps,prb,preprint,groupedaddress]{revtex4}

\usepackage{amsmath}
\usepackage{graphicx}
\usepackage{dcolumn}
\usepackage{bm}
\usepackage{subfigure}

\begin{document}

\title{\emph{Ab initio} Melting Curve of Molybdenum by the Phase Coexistence Method}

\author{C. Cazorla$^{1,2}$}
\author{M. J. Gillan$^{1,2}$}
\author{S. Taioli$^{3}$}
\author{D. Alf\`e$^{1,2,3}$}
\affiliation{$^{1}$London Centre for Nanotechnology, UCL, London WC1H OAH, U.K. \\ 
$^{2}$Department of Physics and Astronomy, UCL, London WC1E 6BT, U.K. \\
$^{3}$Department of Earth Sciences, UCL, London WC1E 6BT, U.K.}

\begin{abstract}
We report \emph{ab initio} calculations of the melting curve of molybdenum for
the pressure range $0 - 400$~GPa. The calculations employ density functional theory
(DFT) with the Perdew-Burke-Ernzerhof exchange-correlation functional in
the projector augmented wave (PAW) implementation. We present tests showing that 
these techniques accurately reproduce experimental data on 
low-temperature b.c.c. Mo, and that PAW
agrees closely with results from the full-potential linearized augmented
plane-wave implementation. The work attempts to overcome the uncertainties inherent
in earlier DFT calculations of the melting curve of Mo, by using the ``reference
coexistence'' technique to determine the melting curve. In this technique,
an empirical reference model (here, the embedded-atom model) is accurately
fitted to DFT molecular dynamics data on the liquid and the high-temperature
solid, the melting curve of the reference model is determined by simulations
of coexisting solid and liquid, and the \emph{ab initio} melting curve is
obtained by applying free-energy corrections. Our calculated melting curve
agrees well with experiment at ambient pressure and is consistent with shock
data at high pressure, but does not agree with the high pressure melting curve
deduced from static compression experiments. Calculated results
for the radial distribution function show that the short-range atomic order of the
liquid is very similar to that of the high-$T$ solid, with a slight \emph{decrease}
of coordination number on passing from solid to liquid. The electronic densities
of states in the two phases show only small differences. The results do not
support a recent theory according to which very low $d T_{\rm m} / d P$
values are expected for b.c.c. transition metals because of electron
redistribution between s-p and d states.
\end{abstract}

\maketitle

\section{Introduction}
\label{sec:introduction}

Over the past five years, a major controversy has developed about the high-pressure melting
curves of transition metals. According to static compression experiments, performed using
diamond anvil cells (DAC) at pressures from ambient up to $\sim 100$~GPa (1~Mbar), the melting temperatures
$T_{\rm m}$ of many transition metals, particularly those having the b.c.c. crystal
structure, change by no more than a few hundred~K over this pressure
range~\cite{errandonea01,errandonea03}. In some 
cases, this finding seems to be in gross conflict with
shock experiments, which indicate an increase of $T_{\rm m}$ of several thousand K
over the same pressure range~\cite{brown83,hixson89,hixson92}. 
Theoretical work~\cite{moriarty94,moriarty02,wang02,belonoshko04} based directly or indirectly on
density functional theory (DFT) generally supports the shock data. We report here
a detailed DFT study of the high-pressure melting curve of Mo, a metal for which
there are some of the largest differences between DAC measurements and other 
data (see e.g. Fig.~2 in Ref.~\cite{errandonea05}).

There has already been quite extensive theoretical work on the high-$P$ melting of metals.
Some of this has been motivated by the desire to understand the properties of
solid and liquid Fe in the Earth's core~\cite{gillan06}. DFT-based calculations of the Fe melting
curve~\cite{alfe99,laio00,belonoshko00,alfe02a} 
up the pressure at the boundary between the solid inner core and the liquid
outer core provide one of the important ways of constraining the temperature
distribution in the core. Disagreements with DAC measurements on Fe and other
transition metals are cause for concern, because if DFT were shown to be
seriously in error, the reliability of DFT for the study of planetary interiors
would be called into question. But even without this practical motivation, a
major disagreement between DFT predictions and experimental data must be taken
seriously, because it suggests an unexpected failure either of commonly used DFT
approximations, or of apparently well established experimental
techniques. Reassuringly, DFT melting curves agree very closely with
experiment for some metals, including Al~\cite{vocadlo02,alfe03} and 
Cu~\cite{vocadlo04}. The large disagreements
arise mainly for transition metals, and the suggestion is that they are linked
to d-band bonding.

Since the melting slope $d T_{\rm m} / d P$ is equal by the Clausius-Clapeyron
relation to $V^{l s} / S^{l s}$, where $V^{l s}$ and $S^{l s}$ are the
volume and entropy of fusion, and since $S^{l s}$ is unlikely to have
exceptionally large values, a very low $d T_{\rm m} / d P$ is likely to
be due to a low volume of fusion. It has been argued~\cite{wittenberg72,errandonea01} that a low
$V^{l s}$ might be expected for b.c.c. transition metals, because the liquid
may be more close packed than the solid. But recently, an additional
argument has been advanced for low $d T_{\rm m} / d P$ values~\cite{ross04}. This
argument depends on the fact that the distribution of conduction electrons
between s-p states and d states is known to depend on the degree of compression
and on the crystal stucture~\cite{pettifor77,pettifor87}. The suggestion is that the (assumed) change of
coordination on going from b.c.c. solid to liquid leads to a change of
electronic structure, and hence a change in the electron distribution
between s-p and d states, and that this redistribution stabilises the
liquid and lowers $T_{\rm m}$. One of the important purposes of the present
work is to use DFT molecular dynamics (m.d.) simulations to test
these suggestions for the case of Mo.

There has been previous DFT work~\cite{moriarty94,belonoshko04} on the high-pressure melting of Mo. 
The early work of Moriarty~\cite{moriarty94} employed a many-body total energy function derived from
first-principles ``generalized pseudopotential theory''~\cite{moriarty90}.
This approximates the total energy of the
system in terms of volume dependent 1-, 2-, 3- and 4-body interatomic
potentials, and accounts for the angular forces that are known to be
important in transition metals. This form of total-energy function is designed
to be fully transferable between different structures, and should be valid
for both the solid and the liquid state.
The model total-energy function was used in m.d. simulations to determine the melting curve by two methods. 
The first method consisted of cycling the simulated system up and down through
the melting point at fixed volume. The results were cross-checked against
a second method based on the calculation of the free energies of the solid
and the liquid. This was pioneering work, but its quantitative accuracy can
be questioned, because some of the phonon frequencies predicted by the total
energy function agreed rather poorly with experiment. In addition, the
surprising claim was made that $T_{\rm m}$ could be changed by up to
a factor of two by the inclusion of thermal electronic excitations, which
were treated only crudely. The predicted
melting curve was consistent with shock data, but was far above the curve
given by recent DAC measurements: the difference of $T_{\rm m}$ values amounts
to $\sim 3000$~K at $P = 100$~GPa. Much more recently, direct DFT m.d. simulation
has been used by Belonoshko {\em et al.}~\cite{belonoshko04} to map out the Mo melting curve
up to $P \simeq 330$~GPa. Their melting curve, like that of Moriarty~\cite{moriarty94}, is consistent
with shock data, but disagrees strongly with DAC data. However, even though
the simulations employed an implementation of DFT that is expected to
be accurate, the method used to map the melting curve may still give
inaccurate results, as noted by the authors. Their method was to heat the
simulated solid at constant volume until the internal energy, pressure,
radial distribution function and self-diffusion coefficient showed
discontinuities attributable to melting. This approach gives an upper
bound to $T_{\rm m}$, but may overestimate it significantly, because of
superheating. The authors estimated that the error in $T_{\rm m}$ due to
superheating should be $\sim 20$~\%, but the arguments for this
estimate are indirect. 

The present new work on Mo melting has several aims. The first aim is
to calculate more accurately the melting curve that follows from the adopted
DFT exchange-correlation functional $E_{\rm xc}$. (We present tests showing that
the GGA-PBE functional is a good choice.) It is only by doing this that the
possible reasons for the disagreement between theory and experiment
can be narrowed down. We have shown in our recent work on other materials
that the errors in computed DFT $T_{\rm m}$ values can be reduced to a few
percent, and we have described several techniques for doing this~\cite{gillan06}. The technique
used here is the ``reference coexistence method''~\cite{alfe02b}. This requires the accurate
fitting of an empirical ``reference'' total energy function to DFT m.d.
simulations on the solid and the liquid; the melting curve of the
reference model is then calculated from simulations on large
systems consisting of coexisting solid and liquid; as an essential
last step, free-energy corrections for the difference between the
reference and DFT total-energy functions are used to correct the
melting curve. This technique was successfully used in our work on
the melting of Cu~\cite{vocadlo04}, and it has been shown to give results
in excellent agreement with an alternative technique, in which DFT
free energies of the solid and liquid are calculated~\cite{alfe99,alfe02b}. Since it
is a `thermodynamic' technique relying on equality of Gibbs free energies of
the two phases, it cannot suffer from superheating problems.
A second important aim is to study the differences of atomic and
electronic structure of the coexisting solid and liquid, to provide
an improved understanding of the factors that determine the melting
curve of Mo. The empirical reference model used to determine the
DFT melting curve has the form of the embedded atom model (EAM)~\cite{daw84,finnis84}.
A useful side benefit of the work is that we obtained a parameterized
EAM that mimics quite well the DFT total-energy function of high-$T$
solid and liquid Mo. This model reveals important features of
the energetics of Mo at high-$P$ and high-$T$, and 
we expect it to be useful in future modelling work
on this metal. 

The remainder of the paper is organized as follows. A brief summary of
the DFT modelling techniques is given in Sec.~\ref{sec:techniques},
followed by an outline of the reference coexistence technique. Then,
Sec.~\ref{sec:tests} details the rather extensive tests we have performed
to ensure that the techniques deliver an accurate description of
the energetics and the vibrational and electronic properties of Mo.
Sec.~\ref{sec:melting} presents our results for the
DFT melting curve of Mo up to 400~GPa, together with the entropy
and volume of fusion as a function of pressure; our comparison
of the atomic and electronic structures of the high-$T$ b.c.c. 
solid and the liquid is reported at the end of the Section. A discussion
of all the results and their relation with previous work is given
in Sec.~\ref{sec:discussion}, followed by our conclusions.

\section{Techniques}
\label{sec:techniques}

\subsection{DFT methods}
\label{sec:DFT_methods}

A comprehensive description of DFT methods as applied to the
modelling of condensed matter is given in recent books~\cite{martin04,kohanoff06}.
As is well known, there is one and only one uncontrollable
approximation in DFT, namely the approximation
used for the exchange-correlation functional $E_{\rm xc}$.
There is abundant evidence that commonly used $E_{\rm xc}$
functionals yield accurate results for a range of properties
of transition-metal crystals, including equilibrium
lattice parameter, elastic constants, phonon frequencies,
$T = 0$ equation of state, solid-state phase
boundaries, etc~\cite{gillan06}. DFT provides the basis for
our present understanding of the electronic structure
of transition-metal crystals as a function of pressure.
There is also some evidence that it accurately reproduces
the structure of liquid transition metals~\cite{gu04}. Nevertheless,
we considered it necessary to test the accuracy
of different $E_{\rm xc}$ in the case of Mo, as we report
in Sec.~\ref{sec:tests}.

A completely separate issue from the choice of $E_{\rm xc}$
is the {\em implementation} of DFT that is used. This
concerns mainly the way that the electron orbitals
are represented. For simulations of the high-$T$ solid
and the liquid, DFT molecular dynamics (m.d.) must
be used, and for this we use the PAW (projector
augmented wave) technique~\cite{bloechl94,kresse99}, which is generally regarded as
the most accurate for m.d. purposes. The PAW method developed by Bl\"{o}chl~\cite{bloechl94}
efficiently combines some of the features originally devised within  
both linear augmented-plane-wave (LAPW, briefly described below) and pseudopotential approaches. 
In PAW, a linear transformation between the all-electron and pseudized wavefunctions
is defined in terms of all-electron and pseudized partial waves and a set
of projector functions localized on the atoms. There is a simple formal
relationship between PAW and the ultra-soft pseudopotential method of
Vanderbilt~\cite{vanderbilt90}, and computationally the two approaches 
are almost equivalent~\cite{kresse99}.

For m.d., we use the method known as
Born-Oppenheimer dynamics, in which the self-consistent
ground state is recalculated at each new m.d. time-step.
Given the criticisms that are sometimes made of
DFT work on melting properties~\cite{errandonea05}, is important to
be clear that this means the recalculation of
the electronic structure of the entire simulated system,
so that changes of electronic structure on going from solid
to liquid at any degree of compression are fully included.
Although PAW is well suited to m.d. simulation, we
regard it as essential to check its accuracy against still
more accurate implementations of DFT, and to do this
we have compared with the predictions of the FP-LAPW
(full-potential linearized augmented plane wave) technique,
which was developed for high-precision calculations on crystals~\cite{lapw1,lapw2,lapw3}.
In this approach, space is divided into spherical regions
centred on the ions, and the interstitial space between the
spheres. Accurate solution of Schr\"{o}dinger's equation
on a radial grid for each angular momentum within each
sphere allows both valence and core states to be treated to arbitrarily
high precision. The FP-LAPW results to be presented include fully relativistic 
treatment of the core states and scalar relativistic treatment of the states in valence~\cite{koelling77}.
In FP-LAPW calculations on crystalline
Mo, we can ensure that they are fully converged with
respect to all technical parameters, so that all errors
of {\em implementation} are negligible, and the only
approximation is that due to $E_{\rm xc}$ itself. All our
PAW and FP-LAPW calculations were performed using the VASP code~\cite{kresse96}
and the WIEN2k code~\cite{wien2k}, respectively.

Normally, the aim of DFT calculations is to obtain the electronic
ground state for given ionic positions. However, because of the
high $T$ involved, it is essential in the present work to include 
thermal electronic excitations, so that for any given ionic
positions we must determine the orbitals and occupation numbers
that self-consistently minimize the electronic {\em free}
energy, using the $T > 0$ version of DFT originally developed
by Mermin~\cite{mermin65}. We know this is essential, because work on other
transition metals~\cite{alfe01} shows that the electronic specific heat
becomes comparable with the vibrational contribution for
$T \sim 5000$~K. Without this, the free energy difference between
solid and liquid might be seriously in error. All our m.d.
simulations are done in the canonical $( N, V, T )$ ensemble,
with the electronic $T$ set equal to the $T$ of the ensemble.

Values of technical parameters ($k$-point sampling, plane-wave
cut-off, etc.) will be given when we present the
calculations.

\subsection{Reference coexistence methods}
\label{sec:reference_coexistence}

There are three steps in the reference coexistence technique~\cite{alfe02b}. First, an empirical reference model
is fitted to {\em ab initio} simulations of the solid and the liquid at thermodynamic states
close to the expected melting curve. Then, the reference model is used to perform simulations
on large systems in which solid and liquid coexist, so as to obtain points on the melting
curve of the model. Finally, the differences between the reference and {\em ab initio}
total energy functions are used to correct the melting properties of the fitted
model so as to obtain the {\em ab initio} melting properties.

In this work, we have used the embedded-atom model (EAM)~\cite{daw84,finnis84} as the reference model. The total
energy function $U_{\rm ref}$ of this model for a system of $N$ atoms has the form:
\begin{equation}
\label{eq:eam}
U_{\rm ref} ( {\bf r}_1 , {\bf r}_2 , \ldots {\bf r}_N ) = 
\frac{1}{2} \epsilon \sum_{i\neq j} \left( \frac{a}{r_{ij}} \right)^{n} - 
C \epsilon \sum_{i} \bigg[ \sum_{j(\neq i)} \left(
\frac{a}{r_{ij}} \right)^{m} \bigg]^{1/2} .
\end{equation}
where $r_{i j} = | {\bf r}_i - {\bf r}_j |$ is the separation of atoms $i$ and $j$.
The first term on the right represents an inverse-power repulsive pair potential, 
while the second (embedding) term describes
the $d$-band bonding. The model is specified by the characteristic length $a$,
the energy scale $\epsilon$, the dimensionless coefficient $C$ characterizing 
the strength of the embedding energy, and the 
embedding and repulsive exponents $m$ and $n$. We wish $U_{\rm ref} ( {\bf r}_1 , \ldots {\bf r}_N )$ to mimic as
closely as possible the {\em ab initio} total energy function $U_{\rm AI} ( {\bf r}_1 , \ldots {\bf r}_N )$, and
we achieve this by adjusting the EAM parameters.

The method for fitting $U_{\rm ref}$ to $U_{\rm AI}$ is designed so as to minimize the
corrections to the melting curve caused by the difference $\Delta U \equiv U_{\rm AI} - U_{\rm ref}$.
We therefore recall the correction scheme before describing the fitting itself.
For given $P$ and $T$, the difference $G_{\rm AI}^{l s} \equiv G_{\rm AI}^l - G_{\rm AI}^s$ 
between the Gibbs free energies of the {\em ab initio} liquid and solid deviates from
the corresponding difference $G_{\rm ref}^{l s} \equiv G_{\rm ref}^l - G_{\rm ref}^s$ of the 
reference liquid and solid, and we write:
\begin{equation}
\label{eq:gibbsdiff}
G^{ls}_{\rm AI}(P,T)=G^{ls}_{\rm ref}(P,T)+ \Delta G^{ls}(P,T) \; .
\end{equation}
It is the shift $\Delta G^{l s} ( P , T )$ caused by changing the total-energy function from
$U_{\rm ref}$ to $U_{\rm AI}$  that causes the shift of melting temperature at given pressure.
To first order, the latter shift is~\cite{alfe02b}:
\begin{equation}
\label{eq:correctiont}
T'_{\rm m} =\frac{\Delta G^{ls}\left(T_{\rm m}^{\rm ref}\right)}{S^{ls}_{\rm ref}} \; ,
\end{equation}
where $S_{\rm ref}^{l s}$ is the difference between the entropies of liquid and solid (i.e. the
entropy of fusion) of the reference system, and $\Delta G^{l s} ( T_{\rm m}^{\rm ref} )$ is 
$\Delta G^{l s}$ evaluated at the melting temperature of the reference system. 

The shift $\Delta G^{l s}$ is the difference of shifts of Gibbs free energies of liquid and solid
caused by the shift $\Delta U \equiv U_{\rm AI} - U_{\rm ref}$ of total energy function.
We find it convenient to perform our simulations at constant volume and temperature.
Under these conditions, the shift of Helmholtz free energy $\Delta F$ arising from 
$\Delta U$ is given by the well-known expansion:
\begin{equation}
\label{eq:helmholtz_exp}
\Delta F = \langle \Delta U\rangle_{\rm ref} - 
\frac{1}{2} \beta \langle \delta\Delta U^{2}\rangle_{\rm ref} + \cdots \quad , 
\end{equation}
where $\beta\equiv 1/k_{\rm B}T$, $\delta\Delta U\equiv \Delta U - \langle \Delta U\rangle_{\rm ref}$, 
and the averages are taken
in the reference ensemble. From $\Delta F$, we obtain the shift of Gibbs free energy at constant pressure as:
\begin{equation}
\label{eq:isocorisobar}
\Delta G = \Delta F - \frac{1}{2}V\kappa_{T}(\Delta P)^{2} \quad ,
\end{equation}
where $\kappa_{T}$ is the isothermal compressibility and $\Delta P$ is 
the change of pressure when $U_{\rm ref}$ is replaced 
by $U_{AI}$ at constant $V$ and $T$.

In fitting the EAM $U_{\rm ref}$ to the {\em ab initio} $U_{\rm AI}$ so as to minimize the corrections we have
just described, we see from Eq.~(\ref{eq:helmholtz_exp}) and (\ref{eq:isocorisobar}) that the effects of
$\langle \Delta U \rangle_{\rm ref}$, $\langle \delta \Delta U^2 \rangle_{\rm ref}$ and $\Delta P$
all need to be made small. Concerning $\langle \Delta U \rangle_{\rm  ref}$, we note that addition
of a position-independent constant to either $U_{\rm AI}$ or $U_{\rm ref}$ has no effect on
the properties of the solid or the liquid, since it simply redefines the energy zero, so
that a large value of $\langle \Delta U \rangle_{\rm ref}$ is not in itself significant.
However, it is crucially important that $\langle \Delta U \rangle_{\rm ref}$ should have almost
the same value in the solid and the liquid, because otherwise there would be a large
shift of $\Delta G^{l s}$. We also seek to make $\langle \delta \Delta U^2 \rangle_{\rm ref}$
and $\Delta P$ small in both phases. To satisfy all these requirements, we perform long
{\em ab initio} m.d. simulations of the solid and the liquid at thermodynamic states near
the expected melting curve. A large number of statistically independent configurations are then
drawn from both these runs, so that we get a large set of $U_{\rm AI}$ values for a collection of
configurations representative of the solid and the liquid. Our procedure is then
to minimize the mean square fluctuations of $\Delta U$ over this whole set: with
$\overline{\Delta U}$ the mean value of $\Delta U$ over the whole set, we minimize
$\overline{\delta \Delta U^2}$, where $\delta \Delta U \equiv \Delta U - \overline{\Delta U}$.
Note that this $\overline{\delta \Delta U^2}$ is not the same quantity as the
$\langle \delta \Delta U^2 \rangle_{\rm ref}$ appearing in Eq.~(\ref{eq:helmholtz_exp}), because
$\overline{\delta \Delta U^2}$ characterizes the fluctuations of $\Delta U$ over a set of configurations
drawn from both liquid and solid. Minimization of $\overline{\delta \Delta U^2}$ has the
effect of reducing simultaneously the difference of $\langle \Delta U \rangle_{\rm ref}$
between solid and liquid, and of reducing $\langle \delta \Delta U^2 \rangle_{\rm ref}$
in the two phases. In order to reduce also $\Delta P^2$ in the two phases, we add
$\Delta P^2$ into the quantity to be minimized, with a suitable weight. As will be described in
Sec.~\ref{sec:results}, we find it necessary to refit the EAM in this way for different
ranges of pressure along the melting curve. A convenient way of characterizing
the quality of fit of $U_{\rm ref}$ to $U_{\rm AI}$ is obtained by dividing
$\overline{\delta \Delta U^2}$ by the mean square value of the fluctuation
of \emph{ab initio} energy $\delta U_{\rm AI} \equiv U_{\rm AI} - \overline{U_{\rm AI}}$,
where $\overline{U_{\rm AI}}$ is the \emph{ab initio} energy averaged over the
collection of solid and liquid configurations. The dimensionless quantity
characterizing the fit is then $\psi \equiv \left[ \overline{\delta \Delta U^2} /
\overline{\delta U_{\rm AI}^2} \right]^{1/2}$; the smaller the value of $\psi$,
the better is the fit.

Following our previous work~\cite{alfe02b,vocadlo02,vocadlo04}, 
our simulations of coexisting solid and liquid with the reference model
are performed in the $( N, V, E )$ ensemble. 
Starting with a supercell containing the perfect b.c.c. crystal, we thermalize it at a temperature
slightly below the expected melting temperature. 
The system remains in the solid state. The simulation is then
halted, and the positions of the atoms in one half of the supercell are held fixed, 
while the other half is heated 
to a very high temperature (typically eight times the expected melting temperature), so that it melts completely.
With the fixed atoms still fixed, the molten part is then rethermalized to the expected melting temperature.
Finally, the fixed atoms are released, thermal velocities are assigned, and the whole system is allowed to evolve
freely at constant ($N,V,E$) for a long time (normally more than 60 ps), so that the solid and liquid come
into equilibrium. The system is monitored by calculating the average number of particles in slices of the cell
taken parallel to the boundary between the solid and liquid. With this
protocol, there is a certain amount of trial and error to find the overall volume that yields the coexisting
solid and liquid system. (An example of the density profile obtained when the two phases
are in stable coexistence will be given in Sec.~\ref{sec:melting}.)
Our main simulations were done on cells containing 6750 atoms, constructed 
as a $15 \times 15 \times 30$ b.c.c. supercell,  
the long axis being perpendicular to the initial liquid-solid boundary. We tested the adequacy of this system
size by repeating the simulations with larger systems (around 10000 atoms), and found no change in the results.

Finally, our best value of the {\em ab initio} melting temperature is obtained by adding to
the reference melting temperature the correction
$T_{\rm m}^\prime$ given by Eq.~(\ref{eq:correctiont}). The reference entropy of
fusion $S_{\rm ref}^{l s}$ needed in this equation is obtained by performing independent
reference m.d. simulations of the solid and the liquid in the $( N, V, T )$ ensemble
at the temperatures at which we made the coexistence simulations, using at each of these
temperatures the solid and liquid volumes that yield the coexistence pressure. These
simulations give the enthalpy of fusion $H_{\rm ref}^{l s}$, from which we obtain
$S_{\rm ref}^{l s}$ from the relation $H_{\rm ref}^{l s} = T_{\rm m}^{\rm ref} S_{\rm ref}^{l s}$.
To obtain the values of $\Delta G^{l s} ( T_{\rm m}^{\rm ref} )$ needed in Eq.~(\ref{eq:correctiont}),
we again perform reference m.d. simulations of solid and liquid at the coexistence temperatures and
volumes, and calculate $U_{\rm AI}$ for a statistically independent set of configurations
drawn from these simulations. These $U_{\rm AI}$ values must, of course, be fully converged
with respect to system size and ${\bf k}$-point sampling. We find that for a system
of 125 atoms and with $U_{\rm AI}$ calculated with a $2 \times 2 \times 2$ ${\bf k}$-point grid
for both solid and liquid, $U_{\rm AI}$ is converged within $5$~meV/atom. These
$U_{\rm AI}$ values are then used to compute $\langle \Delta U \rangle_{\rm ref}$ and
$\langle \delta \Delta U^2 \rangle_{\rm ref}$. The same calculations yield $\Delta P$ for
solid and liquid, which we use in Eq.~(\ref{eq:isocorisobar}).

\section{Tests of methods: zero-temperature}
\label{sec:tests}

To assess the accuracy of the techniques, we have carried out a number of tests
on the zero-temperature crystal in the pressure range $0 - 300$~GPa.
There are three important questions to analyze. First, we want to test the 
accuracy of different exchange-correlation functionals compared with 
experimental data. Second, we aim to study the effect on PAW results of including 
different electronic states in the valence set.
Third, FP-LAPW calculations are performed to 
assess errors incurred by the PAW approximation.
In addition, we have done tests to demonstrate that PAW correctly reproduces 
the changes of electronic structure with 
pressure given by the FP-LAPW method.
Tests of phonon frequencies are also carried out, because of their 
relevance to the vibrational free energy of the system.
    
\subsection{PAW and FP-LAPW calculations}

At very high pressures, states that would normally be treated as core states may respond
significantly to compression. In the case of Mo, the $4p$ states lie only $\sim 35$~eV below the
Fermi energy $E_{\rm F}$ and we always include them in the valence set. The $4s$ states lie
considerably deeper at $\sim 61$~eV below $E_{\rm F}$, and we have examined the effect of including them.
We show in Fig.~\ref{fig:vasp-LDA-PBE} PAW results for the pressure as a function 
of volume at $T = 0$~K, with and without
the $4s$ states included in the valence set, compared with experimental data~\cite{hixson92}. 
Results are shown with both LDA (Ceperley-Alder
parameterization~\cite{ceperley80}) and GGA (Perdew-Burke-Ernzerhof form~\cite{perdew96}) exchange-correlation
functionals. All calculations were performed on a primitive b.c.c. cell, 
with a $32 \times 32 \times 32$ ${\bf k}$-point grid
and with energy cut-offs of $224.5$~eV (without $4s$ states) and $287.6$~eV (with $4s$ states);
these settings ensure energy convergence to better than 1~meV/atom.
With LDA, inclusion of $4 s$ states makes a significant difference, and improves the agreement
with experiment at high pressure, but with GGA the effect of including $4 s$ states is
very small. All the approximations deviate noticeably from experiment. At low pressures,
LDA underestimates the volume by about $3.1$~\%, while GGA overestimates it by $1.2$~\%.
At high pressures, $P \sim 300$~GPa, LDA volumes are only $0.6$~\% below measured values, while
GGA continues to overestimate them by about $1.7$~\%.

To check the accuracy of PAW itself, we have compared with FP-LAPW calculations of the
$T = 0$~K $P ( V )$ relation, performed with the WIEN2k code~\cite{wien2k}.
Our FP-LAPW calculations include fully relativistic effects of the core states
and scalar relativistic treatment of the states in valence~\cite{koelling77}, and the tolerances are chosen so
as to ensure energy convergence to better than 1~meV/atom~\cite{wien2k_tolerances}. 
As shown in Fig.~\ref{fig:wien2kvasp},
the FP-LAPW $P ( V )$ results are almost indistinguishable from the PAW results over
the entire pressure range of interest, with both LDA and GGA functionals. This provides
valuable confirmation of the accuracy of the PAW technique, on which all our present calculations
of melting properties are based. Out of interest, we also performed FP-LAPW calculations
using the recently developed Wu-Cohen form of GGA~\cite{wu06}, which has been reported to give
improved predictions of condensed-matter properties. 
This functional satisfies the same constraints used to construct the PBE functional, but the 
enhancement factor entering the exchange energy density is chosen to match closely the gradient
expansion of Svendsen and von Barth~\cite{svendsen96} for systems having a slowly varying density. 
For this reason, this functional is expected to perform well for solids but not so well 
for atoms or molecules,
where the variation of the valence electron density is generally larger.   
We find that this functional is in much better agreement
with the experimental $P ( V )$ data than either LDA or GGA(PBE). Unfortunately, at the time
most of the present work was done, we had no way of performing Wu-Cohen calculations within
PAW, so we were unable to do melting calculations with this new functional.

\subsection{Electronic density of states}
\label{sec:electronic_states}

The electronic structure of transition metals has been intensively studied, and changes of electronic
structure with increasing pressure have been thoroughly investigated 
both experimentally and theoretically~\cite{pettifor77,pettifor87,wood62,miedema80}. The energetics
of these metals is, of course, strongly dominated by the d-bands, but the s-p bands also play
an important role. An effect that has been well recognised for many years is the pressure
induced upward shift
of the s-p band relative to the d band, caused by the greater spatial extent 
of the s-p orbitals~\cite{pettifor77}.
This relative shift means that increasing pressure causes a transfer of electrons from s-p
to d bands, resulting in an increase of occupancy of the latter. It 
has been proposed recently~\cite{errandonea05, ross04}
that the large differences between theoretical predictions of melting curves and the
results of static compression experiments for some transition metals may be due to an
incorrect treatment of s-p~$\rightarrow$~d transfer in the DFT-based simulations. 
In order to demonstrate that our PAW calculations reproduce the electronic structure
as given by the most accurate DFT methods, we have made detailed comparisons
of the electronic density of
states (DOS) calculated with PAW and FP-LAPW over the pressure range $0 - 300$~GPa.

We show in Fig.~\ref{fig:doswien2kvasp} the electronic DOS from FP-LAPW and PAW
calculated at 0, 150 and 300~GPa, using the GGA(PBE) functional. We note the 
essentially perfect agreement between the two methods. At all pressures, the DOS consists of
two separate parts: a narrow peak at about 35~eV below the Fermi energy, corresponding
to 4p states, and a much broader distribution consisting of the multiple peaks
due to the 4d bands superimposed on the slowly varying DOS of the s-p bands. We note
the characteristic feature of b.c.c. transition metals that the Fermi energy falls
in a deep minimum in the d-band DOS. As expected, the widths of the 4p and 4d parts
of the DOS broaden markedly with increasing pressure. The pressure induced relative
shift of s-p and d bands cannot be clearly seen from the DOS itself. However, it
is very clear from the band structure, shown at pressures of 0 and $300$~GPa in
Fig.~\ref{fig:bandstructure}. The state lying $\sim 7$~eV below the Fermi energy
is the bottom of the s-p band, which lies well below the d states at zero
pressure, but well above the lowest d states at $P = 300$~GPa. We have checked that this relative
shift is precisely reproduced by the PAW calculations. This leaves little doubt that
the PAW techniques, on which all our melting calculations are based, correctly
reproduce this important feature of the pressure dependent electronic structure. 
In Sec.~\ref{sec:melting}, we will present results on the temperature dependence of
the electronic DOS in the solid and the liquid.

\subsection{Phonon frequencies}

The calculation of phonon frequencies is an important test of DFT approximations, because of the
detailed comparisons with experimental data that can be made. It is particularly important
in the context of melting calculations, because, for the harmonic solid, errors in phonon
frequencies translate directly into free energy errors, which are linked with
errors in melting temperature. We present here our calculations of the phonon
dispersion relations of Mo at its experimental volume, using PAW with GGA(PBE)
exchange-correlation. 

The technique for calculating the phonon frequencies is the small
displacement method, as implemented in our {\sc phon} code~\cite{dariophonon}, which we used in 
earlier work on Fe, Al and Cu. In this method,
the elements of the force-constant matrix are obtained by displacing atoms from the perfect-lattice
positions and computing by DFT the forces on all the atoms. For a b.c.c. crystal, it suffices
to displace a single atom along the (111) direction. Since all the calculations
employ periodic boundary conditions, the displaced atom is at the centre of a periodically
repeated supercell. To obtain accurate dispersion relations over the whole Brillouin zone,
this supercell must be large enough so that the elements of the force-constant matrix
have negligible values at its boundaries. In addition, the calculation of the forces
must be converged with respect to electronic ${\bf k}$-point sampling, and to enhance
this convergence we employ Fermi smearing. This smearing itself incurs errors, which
need to be made negligible. In summary, the phonon frequencies must be converged with
respect to atom displacement, supercell size, ${\bf k}$-point sampling and Fermi smearing.
Rather than insisting that every single frequency be converged, it is more
convenient to require that the geometric mean frequency $\bar{\omega}$ be converged.
This quantity is defined by the equation:
\begin{equation}
\label{eq:omega}
\ln \, \bar{\omega} = \frac{1}{N_{\bf{q}i}}\sum_{{\bf q},i}  \ln \left(\omega_{{\bf q}i}\right) , 
\end{equation}
where $\omega_{{\bf q}i}$ is the phonon frequency of branch $i$ at wave vector $\bf q$,
and $N_{\bf{q}i}$ is the number of branches times total number of
{\bf q} points in the sum. It is useful to work with $\bar{\omega}$,
because it is directly related to the harmonic free energy of lattice vibrations, which, well above
the Debye temperature, is equal to $3 k_{\rm B} T \ln ( \hbar \bar{\omega} / k_{\rm B} T )$ per atom.

Our aim is to have $\bar{\omega}$ converged to better than $1$~\% with respect to all technical
parameters. Tests on small supercells show that an atomic displacement of $0.0062$~\AA\ is small
enough to ensure that anharmonic errors are well below this tolerance. To test convergence with
respect to ${\bf k}$-point sampling and Fermi smearing width $\sigma$, we have done extensive
tests on a $2 \times 2 \times 2$ supercell containing 8 atoms (see Table I).
These tests show that for $\sigma = 0.7$~eV, it is easy to achieve excellent ${\bf k}$-point
convergence. But repetition of this at $\sigma = 0.5$~eV reveals that this reduction of $\sigma$
causes $\bar{\omega}$ to change by $\sim 1$~\%. However, further reduction of $\sigma$ to $0.3$~eV
changes $\bar{\omega}$ by less than $0.5$~\%, so that $\bar{\omega}$ appears to be adequately
converged with respect to ${\bf k}$-points and $\sigma$ with a $12 \times 12 \times 12$ ${\bf k}$-point
grid and $\sigma = 0.5$~eV. We then seek convergence with respect to supercell size using this
value of $\sigma$ and a ${\bf k}$-point density that is reduced in inverse proportion to 
supercell size. The results indicate that convergence to our required tolerance is achieved
with the $4 \times 4 \times 4$ supercell of 64 atoms.

Fig.~\ref{fig:vaspphonon} shows a comparison with experimental data of our calculated
phonon frequencies obtained with the $6 \times 6 \times 6$ supercell of $216$ atoms, 
a $4 \times 4 \times 4$ ${\bf k}$-point grid, and $\sigma = 0.5$~eV. According to
our convergence tests, any discrepancy with experimental frequencies of over
1~\% represents a genuine disagreement. The agreement is actually very satisfactory
over most of the Brillouin zone, with typical discrepancies being $1 - 2$~\%. However,
there is a region around the H point ${\bf k} = ( 2 \pi / a_0 ) ( 1, 0, 0 )$, where
there are discrepancies of $\sim 5$~\%. This same H-point problem has been noted
by previous authors who dealt with transition metals (Mo and Nb) and 
used pseudopotential-based methods~\cite{ho82,ho84}.
The origin of this sharp dip in the phonon dispersion curve of Mo at point H has been
related to the nesting of electronic states near the Fermi level~\cite{varma77, varma79} (see figures in Sec.~\ref{sec:electronic_states}), 
so it is likely that by reducing the Fermi smearing and increasing the number of ${\bf k}$-points  
the agreement with experiments would be improved.   
In any case, we believe that since the discrepancies are rather localized in ${\bf k}$-space, they will
have only a weak effect on the thermodynamic properties of the system.

\subsection{Conclusions from the tests}

In summary, our tests show that: (i)~neither LDA nor GGA perfectly reproduces
the experimental $T = 0$~K pressure-volume curve, but the volume given by GGA deviates by
only a small and almost constant amount of $\sim 1.5$~\% from the experimental value over the pressure
range $0 - 300$~GPa; (ii)~with GGA, the inclusion of 4s states in the
valence set makes a negligible difference to the $P ( V )$ curve; (iii)~comparisons of 
PAW and FP-LAPW confirm the accuracy of PAW
for both $P ( V )$ and the electronic DOS, and in particular PAW accurately reproduces
the well-known pressure induced shift of s-p bands relative to d bands; (iv)~GGA
gives rather accurate phonon frequencies over most of the Brillouin zone. This evidence
provides a firm basis for our calculations on the high-pressure melting of Mo,
which employ the PAW technique with GGA(PBE) exchange-correlation, and with
4p states but not 4s states in the valence set.

\section{Melting}
\label{sec:melting}

We begin this Section by presenting our \emph{ab initio} calculations of the
melting curve of Mo using the reference coexistence technique 
(see Sec.~\ref{sec:reference_coexistence}); our results for the
volume and entropy changes on melting as a function of pressure
are also reported. In Sec.~\ref{sec:structure_changes},
we outline our calculations of the atomic and electronic structures
of the b.c.c. solid and the liquid.

\subsection{Calculation of melting curve}
\label{sec:results}

We start by determining the {\em ab initio} $T_{\rm m}$ at a pressure close to zero. At this pressure,
we have experimental values for $T_{\rm m}$ (2883~K)~\cite{shaner77} 
and the volumes per atom of coexisting solid and liquid
(16.34 and 17.04 \AA/atom)~\cite{shaner77}. We use this information to set the temperature and volume/atom of the
{\em ab initio} m.d. simulations that we performed to fit the parameters of the reference model.
These m.d. simulations employed systems of 125~atoms with $\Gamma$-point sampling, and were done
at $T = 2800$~K, $V = 16.34$~\AA$^3$/atom (solid) and $T = 3000$~K, $V = 16.34$~\AA$^3$/atom (liquid). The 
starting configuration of the solid was produced by equilibrating the system, which initially was in the perfect
crystal configuration, to temperature $2800$~K. For the liquid, we produced
the starting configuration by raising the temperature of the perfect crystal to $8000$~K (more than twice the 
experimental $T_{\rm m}$) and then rethermalizing it again to $3000$~K. We checked that 
the system was in the liquid state
by monitoring the time-dependent mean-squared displacement.
The durations of the simulations were about $2$~ps for the solid and $4$~ps for the liquid. 
A set of 100 solid and liquid configurations from these \emph{ab initio}
simulations was then used to fit the reference model by varying the EAM parameters
to minimize the dimensionless quantity $\psi$ and the pressure difference $\Delta P$ 
(see Sec.~\ref{sec:reference_coexistence}).
The resulting EAM parameters are reported in Table II. The very small
value $\psi = 0.078$ indicates a good quality of fit, and the 
$\Delta P$ values of 0.3 GPa and 0.7 GPa for the liquid and solid, respectively, were also satisfactory
(these $\Delta P$ give contributions of only $7$ and $3\cdot 10^{-5}$ eV/atom to $\Delta G$). 
We then performed reference coexistence simulations, and found
that the solid and liquid remain in stable coexistence over periods of $60$~ps at 
$P = 11$~GPa and $T = 3260$~K. To illustrate this, we show in Fig.~\ref{fig:slice}
the density profile obtained by calculating the number of atoms in slices parallel
to the solid-liquid interface. The solid is immediately recognisable from the
regular oscillations with a repeat distance of 2.75~\AA\ (equal to 
$\sqrt{3}/2$ times the lattice parameter of the bcc lattice, corresponding
to the distance between nearest neighbours), whereas the density profile
is flat in the liquid region. Finally, we corrected for the difference between
{\em ab initio} and reference energy functions, obtaining a final {\em ab initio}
$T_{\rm m} = 3205$~K at $P = 11$~GPa. The values of $\langle \Delta U \rangle^{ls}_{\rm ref} / N$
(where $\langle \Delta U \rangle^{ls}_{\rm ref} \equiv \langle \Delta U \rangle^{l}_{\rm ref}  - 
\langle \Delta U \rangle^{s}_{\rm ref}$) and of
$\langle ( \delta \Delta U )^2 \rangle_{\rm ref} / 2 N k_{\rm B} T$ of the liquid and solid used to make these
corrections are reported in Table III. We comment below on the volume and entropy
of melting.

We now use the reference model to obtain a first estimate of the melting curve at higher
pressures. Reference coexistence simulations performed with the EAM parameters
from the fit at $P \simeq 0$~GPa showed that at $P = 92$~GPa, the reference
$T_{\rm m}$ is 5392~K, the volumes of coexisting solid and liquid being
13.30 and 13.53~\AA$^3$/atom. Correcting for the differences between {\em ab initio}
and reference energy functions, we obtain the {\em ab initio} melting temperature
$T_{\rm m} = 4867$~K at $P = 92$~GPa. We note that the correction to $T_{\rm m}$
is considerably greater than at $P \simeq 10$~GPa, but we believe it is still small
enough for the first-order correction scheme to remain valid, and this is confirmed
by subsequent results (see below). However, when we repeated this procedure
at $P \simeq 160$~GPa, still using the reference model
fitted at $P \simeq 0$, we found that the corrections need to go from reference
to {\em ab initio} $T_{\rm m}$ were even larger, and we considered it essential
to refit the reference model. Rather than attempting to do this refit at
$P \simeq 160$~GPa, we returned to $P \simeq 90$~GPa, where our knowledge
of the {\em ab initio} $T_{\rm m}$ is reasonably secure. The refitting at
$P \simeq 90$~GPa produced the new parameters reported in Table II
($100 \le P \le 200$~GPa). This new reference model, when used in coexistence
simulations at $P = 156$~GPa, yielded the reference $T_{\rm m} = 6510$~K,
and a corrected {\em ab initio} value $T_{\rm m} = 5969$~K. We regarded the size of this
correction as acceptably small. In a similar way, when we performed calculations
at $P \simeq 270$~GPa, using the reference model fitted at 90~GPa, the corrections
were unacceptably large, and we performed a refit at 160~GPa. This refitted reference
model required only rather small corrections when used at 270 and 380~GPa.

The reference and {\em ab initio} $T_{\rm m}$ as a function of pressure from
this full set of calculations are reported in Fig.~\ref{fig:melting}. We find
that the {\em ab initio} $T_{\rm m}$ values can be very well fitted with the
so-called Simon equation~\cite{simon29} $T_{\rm m} = a ( 1 + P/b )^c$, with
$a = 2894$~K, $b = 37.22$~GPa and $c = 0.433$. The resulting $P = 0$
melting temperature of 2894~K is very close to the experimental value
of 2883~K. Using the Simon equation, we can obtain the melting slope
$d T_{\rm m} / d P$ at any pressure. At $P = 0$, we find 
$d T_{\rm m} / d P = 33.7$~K~GPa$^{-1}$, which agrees closely with
the experimental value~\cite{shaner77} of $33.3$~K~GPa$^{-1}$. Also shown in Fig.~\ref{fig:melting}
is the point on the melting curve at $P \simeq 375$~GPa estimated from the
shock data of Hixson {\em et al.}~\cite{hixson89}, which is close to our melting curve. The
diamond anvil cell (DAC) measurements of Errandonea {\em et al.}~\cite{errandonea01} differ
greatly from our results, since their $d T_{\rm m} / d P$ is
essentially zero over most the range from 0 to 100~GPa. Previous
theoretical melting curves~\cite{moriarty94,belonoshko04} for Mo, also shown in the Figure, are
in general agreement with our results, though there are substantial
quantitative differences. The comparison of all these experimental
and theoretical results raises important issues, which will be
discussed in Sec.~\ref{sec:discussion}.

The entropy and volume of fusion of the reference model, denoted by $S_{\rm ref}^{l s}$ and $V_{\rm ref}^{l s}$,
are straightforward to calculate. From the reference coexistence simulations, we have $( P, T )$ pairs
lying on the reference melting curve. We then perform separate single-phase m.d. simulations of the
solid and liquid reference systems at chosen temperatures, using systems of 3375 atoms, adjusting
the volumes in each case to give the appropriate $P$. The difference of the resulting volumes give
us $V_{\rm ref}^{l s}$. At the same time, the simulations give the internal energy $U$, from which
we obtain the enthalpies $H = U + P V$ of the two phases. Then the difference of enthalpies
$H_{\rm ref}^{l s}$ gives us the entropy difference, since $T S_{\rm ref}^{l s} = H_{\rm ref}^{l s}$.
We find that $S_{\rm ref}^{l s}$ is almost constant along the reference melting curve, going from
$0.58$~$k_{\rm B}$/atom at $P \sim 0$ to $0.69$~$k_{\rm B}$/atom at $P = 378$~GPa. Since the reference
system mimics the {\em ab initio} system rather closely, we assume that $S_{\rm AI}^{l s}$ 
for the {\em ab initio} system is essentially the same as $S_{\rm ref}^{l s}$. To obtain
the {\em ab initio} volume of fusion $V_{\rm AI}^{l s}$, we use the Clausius-Clapeyron
relation $d T_{\rm m} / d P = V_{\rm AI}^{l s} / S_{\rm AI}^{l s}$. Our $V^{l s}$ results
for the reference and {\em ab initio} systems as a function of $P$ are reported in Fig.~\ref{fig:changevol}.
We note that in both cases the fractional volume change $V^{l s} / V^s$ decreases smoothly
from $\sim 1.5$~\% at $P = 0$ to $\sim 0.9$~\% at 400~GPa.

\subsection{Atomic and electronic structure of solid and liquid}
\label{sec:structure_changes}

We noted in the Introduction that theories of the melting of b.c.c. transition metals sometimes
assume~\cite{ross04} that melting is associated with a significant change of coordination number, so that
the electronic density of states should also change markedly. In order to test 
these ideas for Mo, we have calculated the radial distribution function $g ( r )$
of Mo from a series of AIMD simulations of the solid and the liquid performed
at $P \simeq 216$~GPa. The simulations were all done with 125 atoms, and had
a typical duration of 2~ps after equilibration. Fig.~\ref{fig:gr} reports our calculated
$g ( r )$ at $T = 2000$, 4000 and 6000~K (solid state), and at $T = 7500$~K (liquid state); we
recall (Fig.~\ref{fig:melting}) that our calculated $T_{\rm m}$ at this pressure is $6641$~K.
At 2000~K, the shells containing 8 first neighbours and 6 second neighbours at distances
of $r = 2.44$ and 2.81~\AA\ are clearly separated, and $g ( r )$ goes to zero at $r \simeq 3.4$~\AA,
between the second and third shells. However, at $T = 4000$~K, the first and second shells have already
merged, though the shoulder due to the second shell is clearly visible. The $g ( r )$
of the solid at $T = 6000$~K presents similar trends to those shown at 4000 K, the main difference
being that the valley between the second and third atomic shells now clearly moves upwards from zero.
The change in $g ( r )$ in going from
the solid at $T = 6000$~K to the liquid at $7500$~K is substantial for $r > 3$~\AA, with the
well defined peaks due to third and higher neighbours in the solid becoming heavily
broadened. However, the peak closest to the origin does not suffer a large change. We define
the coordination number $N_{\rm c}$ in the usual way as 
$N_{\rm c} = 4 \pi \bar{\rho} \int_0^{r_{\rm c}} g(r) r^2 \, dr$, with $\bar{\rho}$ the
bulk number density and $r_{\rm c}$ the distance of the first minimum. The $r_{\rm c}$ values
at the four temperatures are 3.41, 3.41, 3.41 and 3.31~\AA, and the resulting $N_{\rm c}$ values 
are 14, 14, 14 and 13.35. Apart from the expected increase in disorder, the main change
on going from solid to liquid is thus a slight {\em decrease} in coordination number. We
comment further on this in Sec.~\ref{sec:discussion}.

Turning now to the electronic density of states (DOS), we present first our AIMD
results for $P$ in the range $50 - 70$~GPa at a series of temperatures, the simulations
being performed on a system of 125 atoms with $\Gamma$-point sampling. The typical
duration of these simulations was 2~ps after equilibration and the DOS were calculated 
by averaging over 150 different configurations. We report in Fig.~\ref{fig:dosmelt}
the calculated DOS at the thermodynamic states given by $( P, T ) = ( 48, 0 )$, $( 50, 2000 )$,
$( 51, 3300 )$ for the solid, and $( P, T ) = ( 72, 5000 )$ for the liquid (units of
GPa and K). (Our {\em ab initio} $T_{\rm m}$ for $50 < P < 70$~GPa are in the range
$4185 - 4577$~K.) We have checked in each case that the system is in the solid or
liquid states by looking at the mean-squared displacement and structure factor.  
We note the progressive broadening of the DOS peaks with 
increasing thermal disorder in the solid, an effect which
continues further in the liquid. The Fermi-level value of the DOS
increases slightly on melting. As far as occupied states are concerned, melting
appears to cause a slight redistribution of d-states from lower in the band to the
region of the Fermi level. 

In the right-hand panel of Fig.~\ref{fig:dosmelt}, we compare our AIMD results
for the electronic DOS at the solid state-point $( P, T ) = (285, 7000 )$ and
the liquid state-point $( P, T ) = ( 300, 8250 )$ (units of GPa and K), which are
just below and just above our calculated melting curve. We note the rather minor
changes caused by melting. Interestingly, the Fermi-level value of the DOS
is almost identical in the two phases at this pressure. The relationship of
these results with earlier work on the electronic structure of liquid
transition metals will be discussed in the following Section.

\section{Discussion and Conclusions}
\label{sec:discussion}

At the start of this paper, we emphasized the large discrepancies between melting curves
of transition metals derived from static compression and shock measurements, and
we mentioned that previous DFT work on Mo supports the shock measurements. The
present work fully confirms that the melting curve predicted by DFT in the
PBE approximation for exchange-correlation energy lies far above the static
compression measurements, but at high pressures is consistent with the shock data. 
This confirmation is important, because of deficiencies or uncertainties in
previous DFT work. The reliability of the present calculations is supported
by our close agreement with the experimental $P = 0$ values of both the melting
temperature $T_{\rm m}$ and the melting slope $d T_{\rm m} / d P$. Our
melting curve is below the theoretical curve of Moriarty~\cite{moriarty94} by $\sim 600$~K at
$P = 0$, and this difference increases with increasing $P$. However, this is
not surprising, since the generalized pseudopotential model that he used is
known to disagree with experimental phonon frequencies, and because
he included thermal electronic excitations only approximately. In the
present work, we have taken pains to verify the accuracy of the phonon
frequencies given by our methods, and thermal electronic excitations
are fully included within our DFT framework. Perhaps more surprising is
that our melting curve agrees closely with that obtained by Belonoshko
{\em et al.}~\cite{belonoshko04} using direct DFT m.d. simulation. This is unexpected,
since they believed that their melting curve suffered from a substantial
superheating error of $\sim 20$~\%. The close agreement suggests that
they may have been unduly pessimistic, and this point deserves further
investigation. We included in Fig.~\ref{fig:melting} the results of
Belonoshko \emph{et al.}~\cite{belonoshko04} and of Verma \emph{et al.}~\cite{verma04}
obtained by the dislocation-mediated theory of melting~\cite{burakovsky00,preston92}.
It is not clear to us whether one can expect a theory of melting based exclusively on the
properties of the solid to be fully reliable. One of the problems with this
approach is that the predicted melting curves rely on thermodynamic data
that may not be reliably known. The rather large differences between the two melting
curves based on the dislocation theory may be indicative of the
limited reliability of this approach.

The change of volume on melting of $\sim 1$~\% given by our calculations is small,
but still much greater than the volume change implied by the static compression
values~\cite{errandonea01,ross04,errandonea05} of $d T_{\rm m} / d P$. Arguments in favour of a very low 
volume change based on a significant increase of coordination
number on going from b.c.c. solid to melt appear to be incorrect, according
to our DFT m.d. calculations of the radial distribution function $g ( r )$.
We find only rather minor differences between $g ( r )$ for high-$T$ solid and
melt. In particular, there is actually a slight {\em decrease} in
coordination number from 14 to $\sim 13$ on melting, so that the liquid
is slightly less close packed than the solid. We comment that ideas
based on hard-sphere packing are likely to be misleading, since the
repulsive interactions between Mo atoms at high $T$ are rather soft (see below).

We mentioned in the Introduction the recent theory of Ross {\em et al.}~\cite{ross04}, according to
which a very low melting slope is expected for b.c.c. transition metals. The theory
invokes the well known transfer of electrons from s-p to d states with increasing
compression, and the fact that this transfer depends on crystal structure.
In applying this theory to the melting of Mo, the authors estimated the
effective number of d electrons $n_{\rm d}$ by treating the high-temperature
solid as a perfect b.c.c. crystal and the liquid as a perfect f.c.c. crystal.
They also assumed that a change of $n_{\rm d}$ on melting will be associated with
a change of d-band width. They found that the changes of electronic structure
stabilize the liquid relative to the solid, and yield a major reduction of $T_{\rm m}$.
In considering this theory in the light of the results we have presented, it is
important to appreciate that our calculations are all based on an accurate
implementation of DFT. As in all simulations using Born-Oppenheimer DFT m.d.,
the VASP code recalculates the entire self-consistent electronic structure at
every time step of the time evolution. As described in Sec.~\ref{sec:electronic_states}, we have gone
to considerable lengths to show that the PAW implementation of DFT used in our m.d. 
yields results for the electronic DOS which are almost indistinguishable
from those given by the FP-LAPW technique, which is one of the most accurate
available. This means that all the effects that enter the theory of Ross {\em et al.}~\cite{ross04},
including s-p to d electron transfer and changes of d-band width, are fully
included in our simulations. Nevertheless, we do not obtain the very low
melting slope that they predict. The reason for this is presumably that
their treatment of the high-$T$ solid and the liquid as perfect crystals
is incorrect. As we have seen, their assumption of a large
structural change on melting also appears to be questionable. This point
is reinforced by our finding that the electronic DOS changes only slightly
on melting, especially at high $P$. 

We end this discussion by commenting on the embedded-atom model (EAM) used as
a reference system in determining the melting curve. It is an important
finding of this work that the EAM is able to mimic accurately the DFT
total-energy function of solid and liquid close to coexistence. This does
not mean, however, that we accept the melting curve of the reference model
as the true melting curve. This could be dangerous, because we
might then miss d-band electronic effects that were not explicitly
included in the model. However, in our procedure, any such effects
are automatically picked up in the corrections that we apply, since these
explicitly account for free energy differences between the reference and
DFT systems. We note in passing that the fitting of our reference model yields
parameters that resemble those we found in our earlier work on the melting
of Fe~\cite{alfe02a}. In particular, the inverse-power repulsive potential in our
present EAM model has an exponent $n$ close to 6 at low $P$, decreasing
to $\sim 5$ at high $P$. For comparison, the fitted EAM in our Fe work
had $n = 5.9$, which is very similar. We are currently investigating
the systematic behaviour of EAM parameters in solid and liquid
transition metals at high $P$ and $T$, and we hope to report on
this elsewhere. 

The main conclusions from this work are as follows. Our DFT calculations of the
melting curve of Mo up to 400~GPa fully confirm earlier work showing that
the DFT melting curve is consistent with shock data, but is far above
the melting curve given by static compression experiments. Our calculations
indicate that at high $P$ there are only minor changes of both atomic and electronic
structure on going from the high-temperature b.c.c. solid to the melt. Suggested mechanisms
for an anomalously low melting slope $d T_{\rm m} / d P$ of Mo based on transfer
of electrons from s-p states to d states appear to be incompatible with the
present DFT calculations. This tends to confirm earlier suggestions~\cite{belonoshko04} that the
transition identified as melting in high-$P$ static compression experiments
may not be true thermodynamic melting. 

\acknowledgments
The work was supported by EPSRC grant EP/C534360, which is 50\% funded
by DSTL(MOD), and by NERC grant NE/C51889X/1. The work was conducted as part
of a EURYI scheme award to DA as provided by EPSRC (see www.esf.org/euryi).

\clearpage

\begin{table}[c]
\begin{center}
\label{tab:omega_convergence}
\begin{tabular}{|c|c|c|c|c|}
\hline
$ N $  &   supercell $ $ &  \textbf{k}-grid $ $ & $ \sigma $ (eV) & $ \overline{\omega}$ ($10^{12}$~s$^{-1}$) \\
\hline
$\quad 8 \quad $  &   $\quad  2\times 2\times 2 \quad $  &   $\quad 4\times 4\times 4 \quad $  &  $\quad 0.7 \quad $  &  $\quad 20.734 \quad $  \\ 
$   $  &   $   $  &   $8\times 8\times 8  $  &      $ 0.7 $  &  $21.385 $  \\ 
$   $  &   $   $  &   $12\times 12\times 12  $  &   $ 0.7 $  &  $21.204 $  \\ 
$   $  &   $   $  &   $16\times 16\times 16  $  &   $ 0.7 $  &  $21.206 $  \\ 
\hline
$ 8  $  &   $ 2\times 2\times 2  $  &   $8\times 8\times 8     $  &   $ 0.5 $  &  $21.498 $  \\ 
$   $  &    $  $ &   $12\times 12\times 12  $  &   $ 0.5 $  &  $21.045 $  \\ 
$   $  &    $  $ &   $16\times 16\times 16  $  &   $ 0.5 $  &  $21.037 $  \\ 
$   $  &    $  $ &   $24\times 24\times 24  $  &   $ 0.5 $  &  $21.045 $  \\ 
\hline
$ 8  $  &   $ 2\times 2\times 2  $  &   $8\times 8\times 8  $     &   $ 0.3  $  &  $21.703 $  \\ 
$   $  &   $  $  &   $12\times 12\times 12  $  &   $ 0.3  $  &  $20.921 $  \\ 
$   $  &   $  $  &   $16\times 16\times 16  $  &   $ 0.3  $  &  $20.811 $  \\ 
$   $  &   $  $  &   $24\times 24\times 24  $  &   $ 0.3  $  &  $20.951 $  \\ 
$   $  &   $  $  &   $32\times 32\times 32  $  &   $ 0.3  $  &  $20.941 $  \\ 
\hline
$ 64 $  &   $ 4\times 4\times 4  $  &   $6\times 6\times 6     $  &   $ 0.5  $  &  $21.699 $  \\ 
$216 $  &   $ 6\times 6\times 6  $  &   $4\times 4\times 4     $  &   $ 0.5  $  &  $21.727 $  \\ 
\hline
\end{tabular}
\end{center}
\caption{ Convergence of mean phonon frequency $\overline{\omega}$ 
          (see Eq.~\ref{eq:omega}) with supercell size, $\textbf{k}$-grid and  
          Fermi broadening $\sigma$. }

\end{table}

\clearpage

\begin{table}[c]
\begin{center}
\label{tab:EAMparameters}
\begin{tabular}{@{\hspace{1.cm}} c@{\hspace{1.cm}}  c@{\hspace{1.cm}}  c@{\hspace{1.cm}}  c@{\hspace{1.cm}}  c@{\hspace{1.cm}}  c@{\hspace{1.cm}} }
\hline
\hline
$ \rm{P} $ (GPa) &  $\epsilon $ (eV)  & $ a $ (\AA) &  $ n $  &  $ m $ &  $ C $  \\
\hline
$ 0-100    $  &   $ 0.323  $    &   $ 3.579  $  &  $ 5.93 $  &  $ 3.72 $  &  $ 12.66  $  \\ 
$ 100-200  $  &   $ 0.169  $    &   $ 4.985  $  &  $ 4.96 $  &  $ 3.88 $  &  $ 28.08  $  \\ 
$ 200-400  $  &   $ 0.144  $    &   $ 4.760  $  &  $ 5.07 $  &  $ 3.78 $  &  $ 26.90  $  \\ 
\hline
\hline
\end{tabular}
\end{center}
\caption{ Parameters of the EAM potential deduced for Mo and used for the coexistence simulations.
          Values are obtained by fitting to \emph{ab initio} simulations on solid and liquid.}
\end{table}

\clearpage

\begin{table}[c]
\begin{center}
\label{tab:gauss}
\begin{tabular}{@{\hspace{1.0cm}} c@{\hspace{1.0cm}}  c  @{\hspace{0.15cm}}c@{\hspace{1.0cm}}  c  @{\hspace{0.15cm}}c @{\hspace{1.0cm}}c@{\hspace{1.0cm}} }
\hline
\hline
$ T_{\rm m}^{\rm ref} $ (K)  &  $ \langle \Delta U\rangle^{ls}_{\rm ref} / N $ (eV/atom) & 
$  $ & $ \frac{1}{2}\beta\langle\left(\delta \Delta U\right)^{2}\rangle_{\rm ref}/N $ (eV/atom) &  
$  $ & $ T_{\rm m}^{\rm AI} $ (K)  \\ \cline{4-4} 
$   $  &   $      $  & $  $ &  $\rm{Solid} {\hspace{1.0cm}} \rm{Liquid}$ & $  $ &  $   $  \\ 
\hline
$  3260  $  &   $ 0.009(2)  $   & $  $  &  $ 0.038(2) {\hspace{0.45cm}} 0.032(2)  $  &   $     $  &  $ 3205 $ \\
$  5392  $  &   $ 0.027(1)  $   & $  $  &  $ 0.024(2) {\hspace{0.45cm}} 0.028(1)  $  &   $     $  &  $ 4867 $ \\
$  6510  $  &   $ 0.038(1)  $   & $  $  &  $ 0.035(3) {\hspace{0.45cm}} 0.030(2)  $  &   $     $  &  $ 5969 $ \\
$  7324  $  &   $ 0.002(1)  $   & $  $  &  $ 0.023(2) {\hspace{0.45cm}} 0.015(2)  $  &   $     $  &  $ 7281 $ \\
$  8618  $  &   $ 0.013(1)  $   & $  $  &  $ 0.018(2) {\hspace{0.45cm}} 0.032(2)  $  &   $     $  &  $ 8154 $ \\
\hline
\hline
\end{tabular}
\end{center}
\caption{Difference $\langle \Delta U \rangle_{\rm ref}^{l s} \equiv \langle \Delta U \rangle_{\rm ref}^l -
          \langle \Delta U \rangle_{\rm ref}^s$ between the liquid and solid thermal averages 
          of the difference
          $\Delta U \equiv U_{\rm AI} - U_{\rm ref}$ of \emph{ab initio} and reference energies,
	  and thermal averages in solid and liquid $\langle \left( \delta \Delta U \right)^2\rangle_{\rm ref}$
	  of the squared fluctuations of 
	  $\delta \Delta U \equiv \Delta U - \langle \Delta U \rangle_{\rm ref}$, 
	  with averages evaluated in the reference 
	  system and normalized by dividing by the number of atoms $N$. 
	  Melting temperatures for the reference and \emph{ab initio} systems 
          are also reported.}
  
\end{table}

\clearpage

{\Huge{\bf Figure Caption List}}

\vspace{1.0cm}

\begin{figure}[c]
\centerline{
        \includegraphics[width=0.8\linewidth]{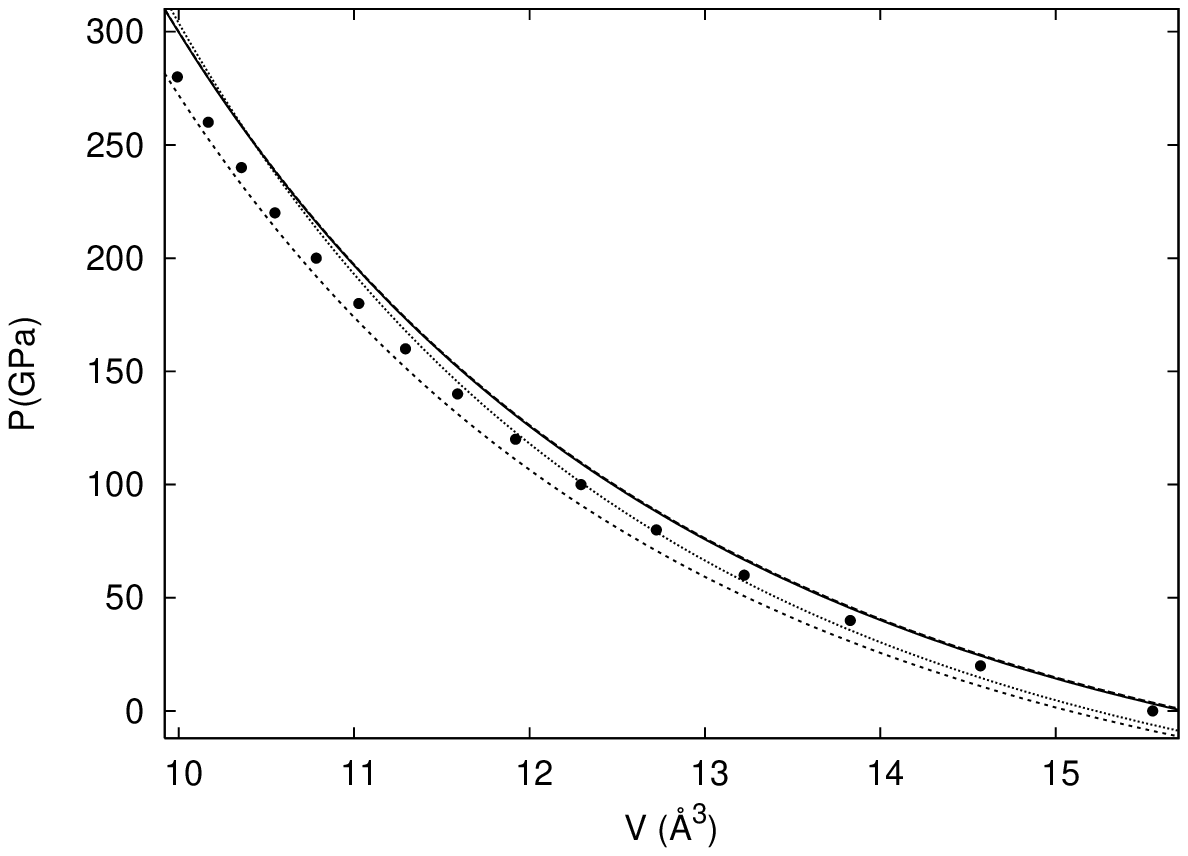}}%
        \caption{}
\label{fig:vasp-LDA-PBE}
\end{figure}

FIG.~\ref{fig:vasp-LDA-PBE}: Comparison of LDA and GGA pressure $P$ as function of
                  volume $V$ for b.c.c. Mo from the PAW method with different
                  exchange-correlation functionals and valence sets.
                  Long-dashed and solid lines (practically coincident) show GGA
                  results with and without $4s$ states in the valence set.
                  Short-dashed and dotted lines show LDA results with and without
                  $4s$ states in the valence set. Dots show experimental results~\cite{hixson92}.\\

\begin{figure}[c]
\centerline{
        \includegraphics[width=0.8\linewidth]{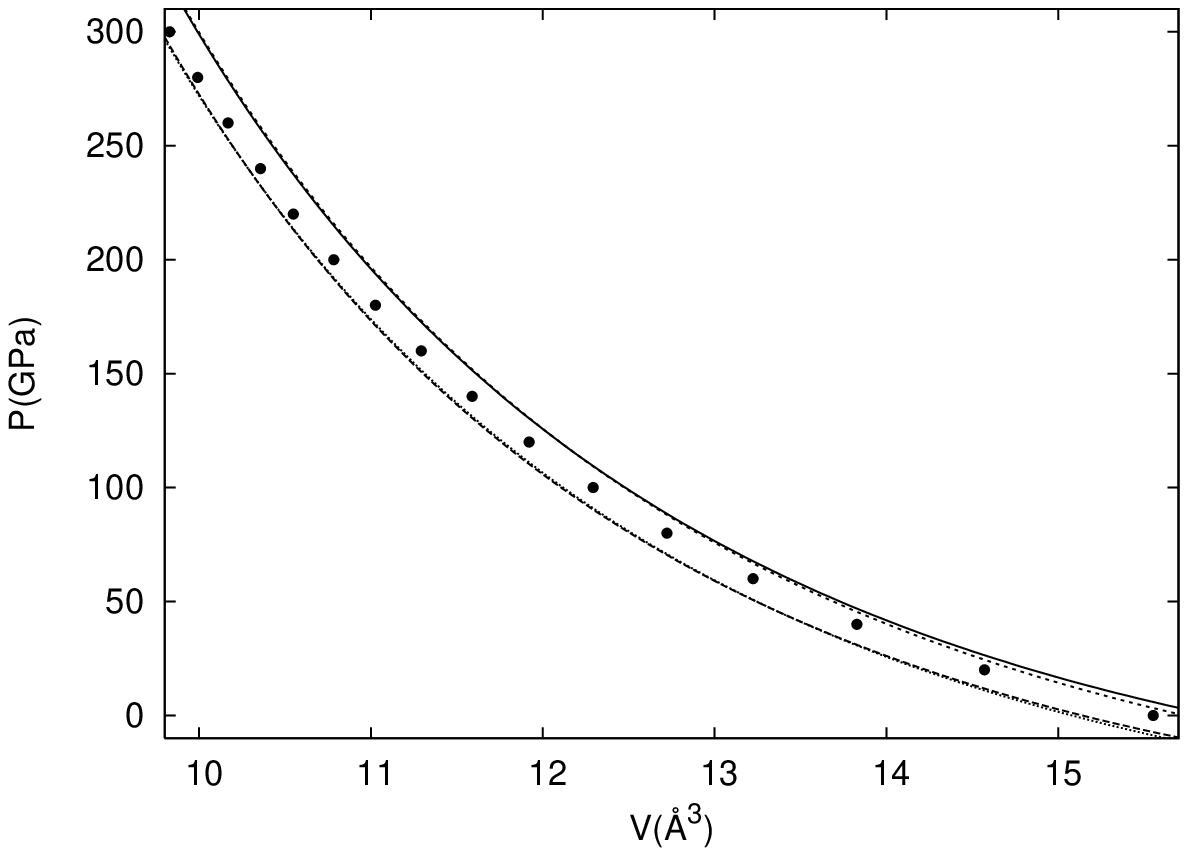}}%
        \caption{}
\label{fig:wien2kvasp}
\end{figure}

FIG.~\ref{fig:wien2kvasp}: Comparison between PAW and FP-LAPW results for the GGA(PBE) and
                  LDA(CA) approximations for $E_{\rm xc}$. Solid and dashed curves show
                  GGA(PBE) and LDA(CA) FP-LAPW results, respectively; short-dashed
                  and dotted curves show GGA(PBE) and LDA(CA) PAW calculations,
                  respectively. Solid dots show experimental data~\cite{hixson92}.\\

\begin{figure}[c]
\centerline{
        \includegraphics[width=0.8\linewidth]{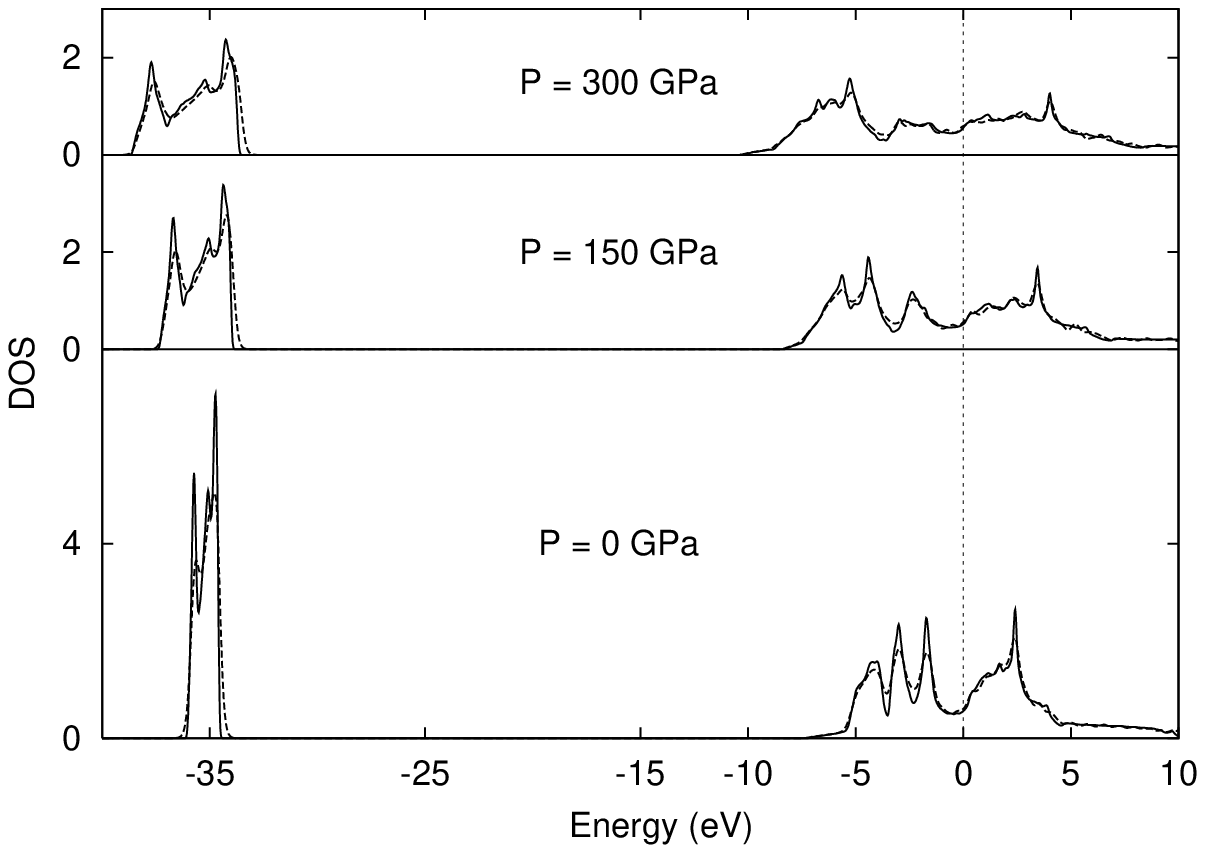}}%
        \caption{}
\label{fig:doswien2kvasp}
\end{figure}

FIG.~\ref{fig:doswien2kvasp}: Density of electronic states obtained with the PAW (dashed line) and FP-LAPW (solid line)
                  at zero temperature and 0, 150 and 300 GPa. Fermi energies are shifted to zero (dotted line).\\

\begin{figure}[c]
\centering
       { \includegraphics[width=0.48\linewidth]{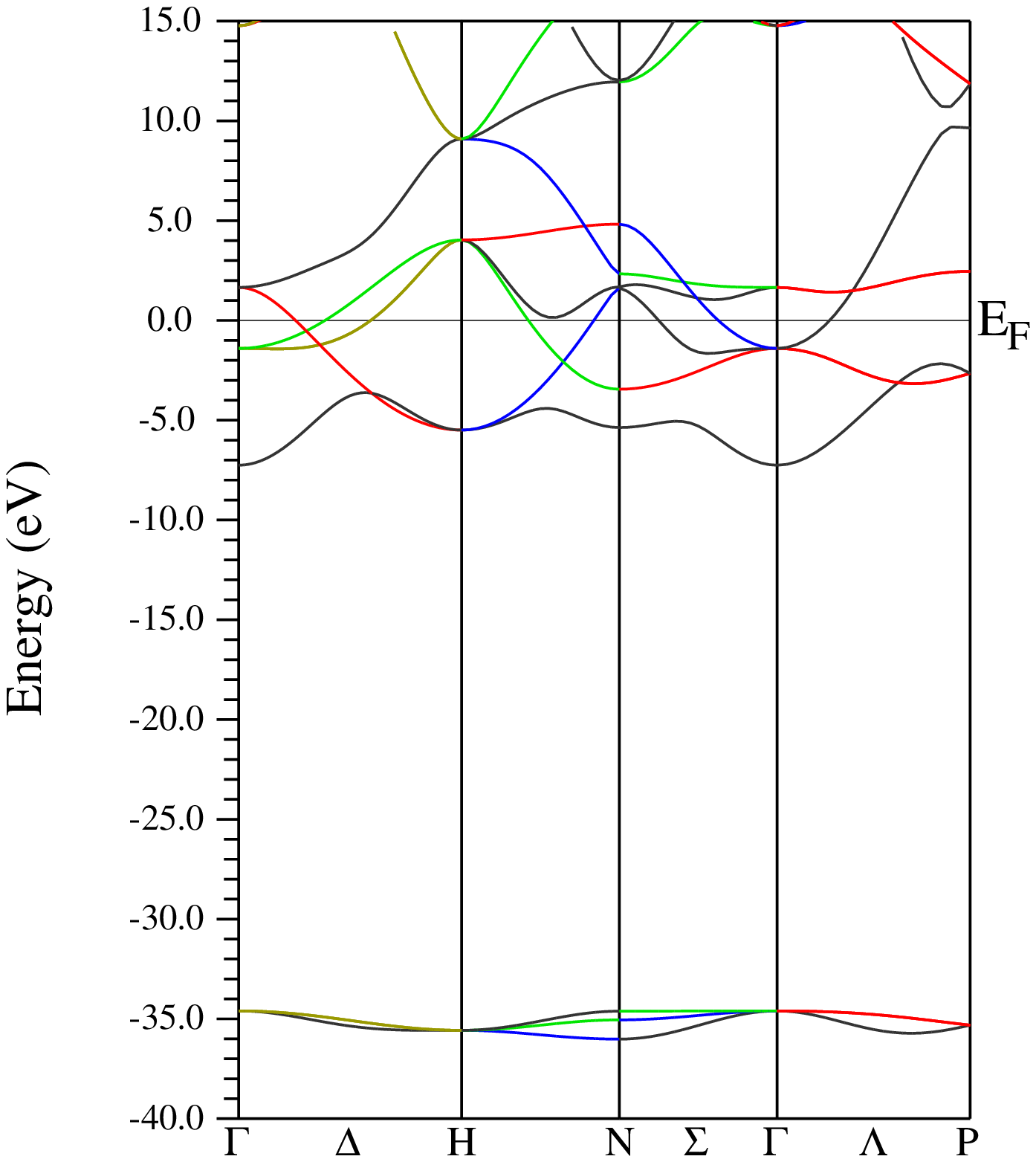} }%
       { \includegraphics[width=0.48\linewidth]{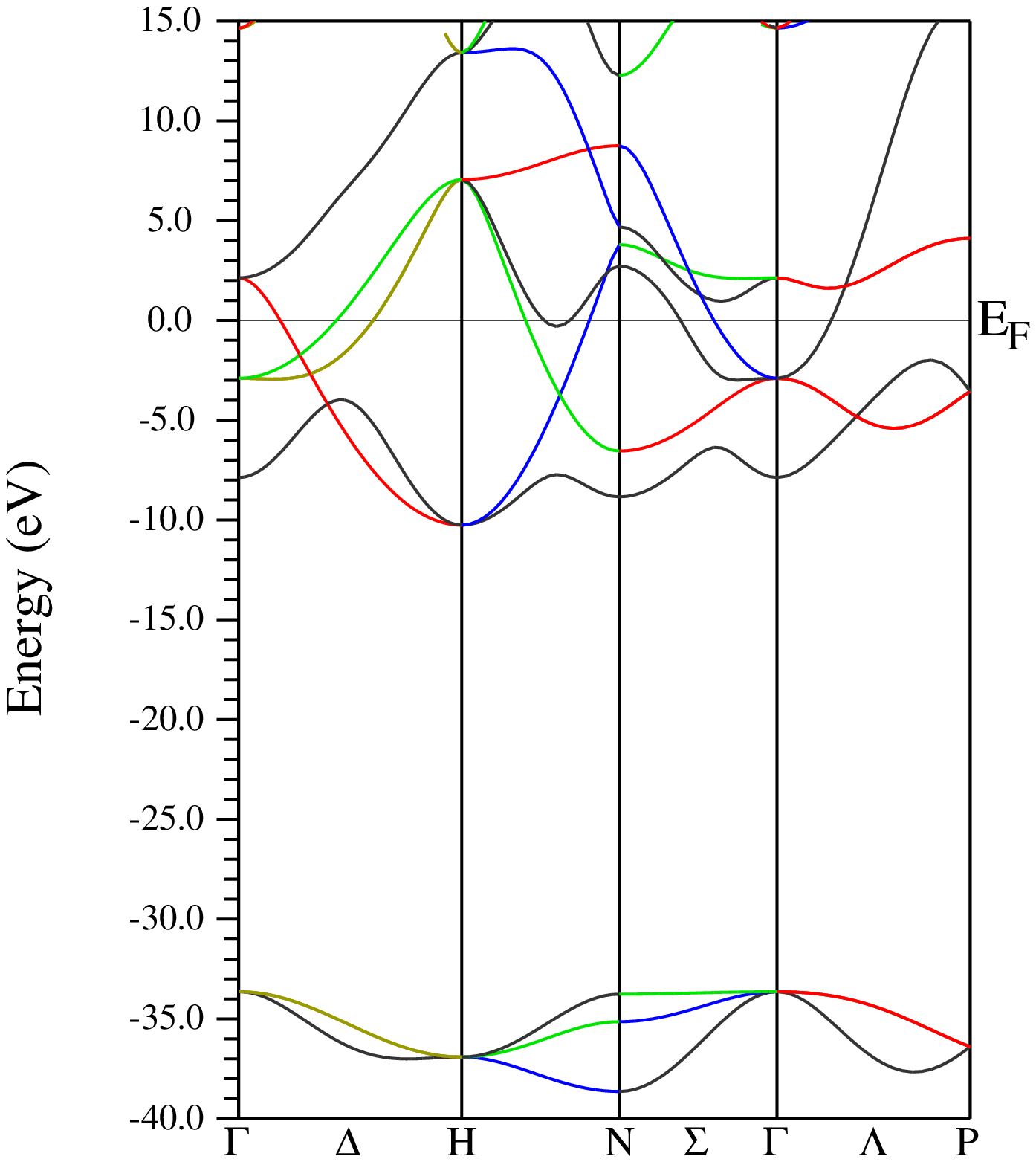} }%
 \vspace{-0.8cm}
 \caption{}
\label{fig:bandstructure}
\end{figure}

FIG.~\ref{fig:bandstructure}: FP-LAPW calculation of the energy bands of Mo at 0 and 300~GPa
          (left and right panels respectively). The $s$ valence band (energy between -8 and -7 eV at the $\Gamma$ point)
          rises in energy more quickly than $d$ valence bands with increasing pressure.\\ 

\begin{figure}[c]
\centerline{
        \includegraphics[width=0.8\linewidth]{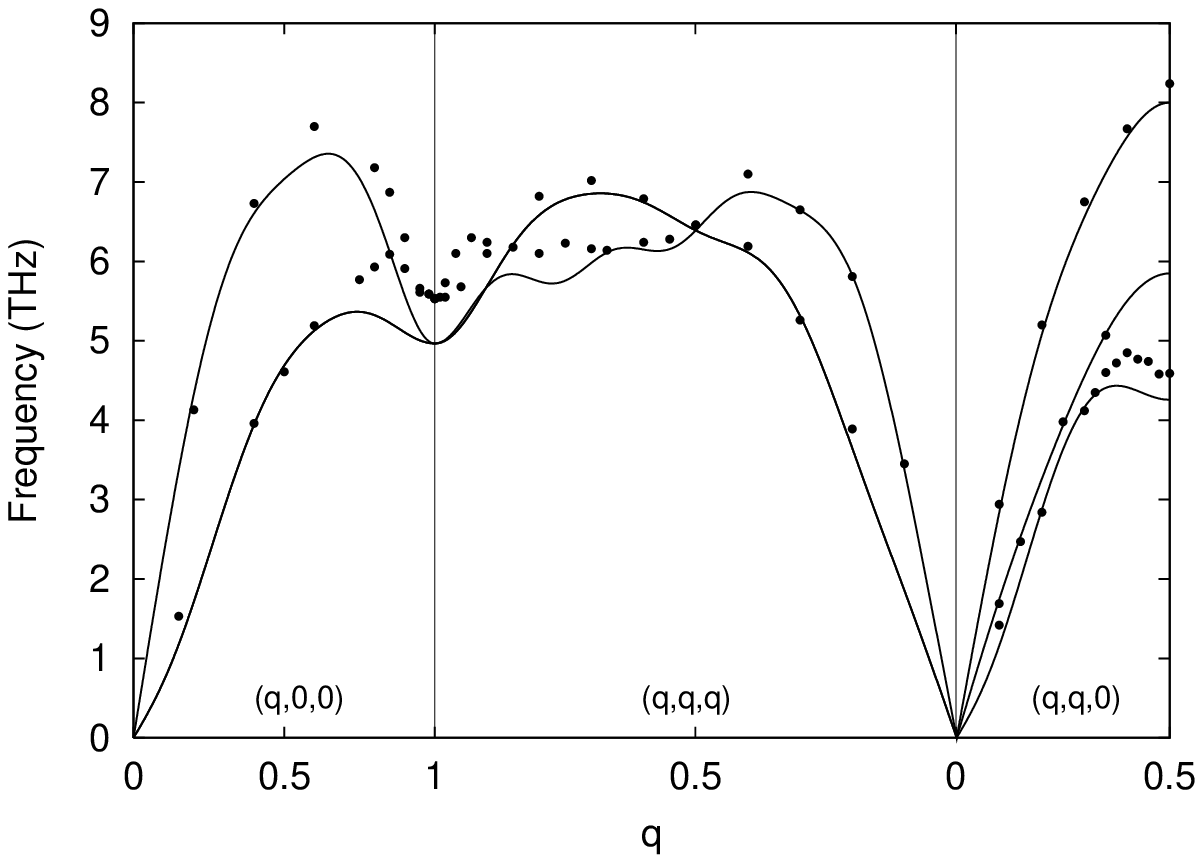}}%
        \caption{}
\label{fig:vaspphonon}
\end{figure}

FIG.~\ref{fig:vaspphonon}: Comparison of calculated (curves) and experimental (solid
        squares) phonon dispersion relations of Mo at zero pressure.
        Experimental data are from Ref.~\cite{zarestky83}.\\

\begin{figure}[c]
\centerline{
        \includegraphics[width=0.8\linewidth]{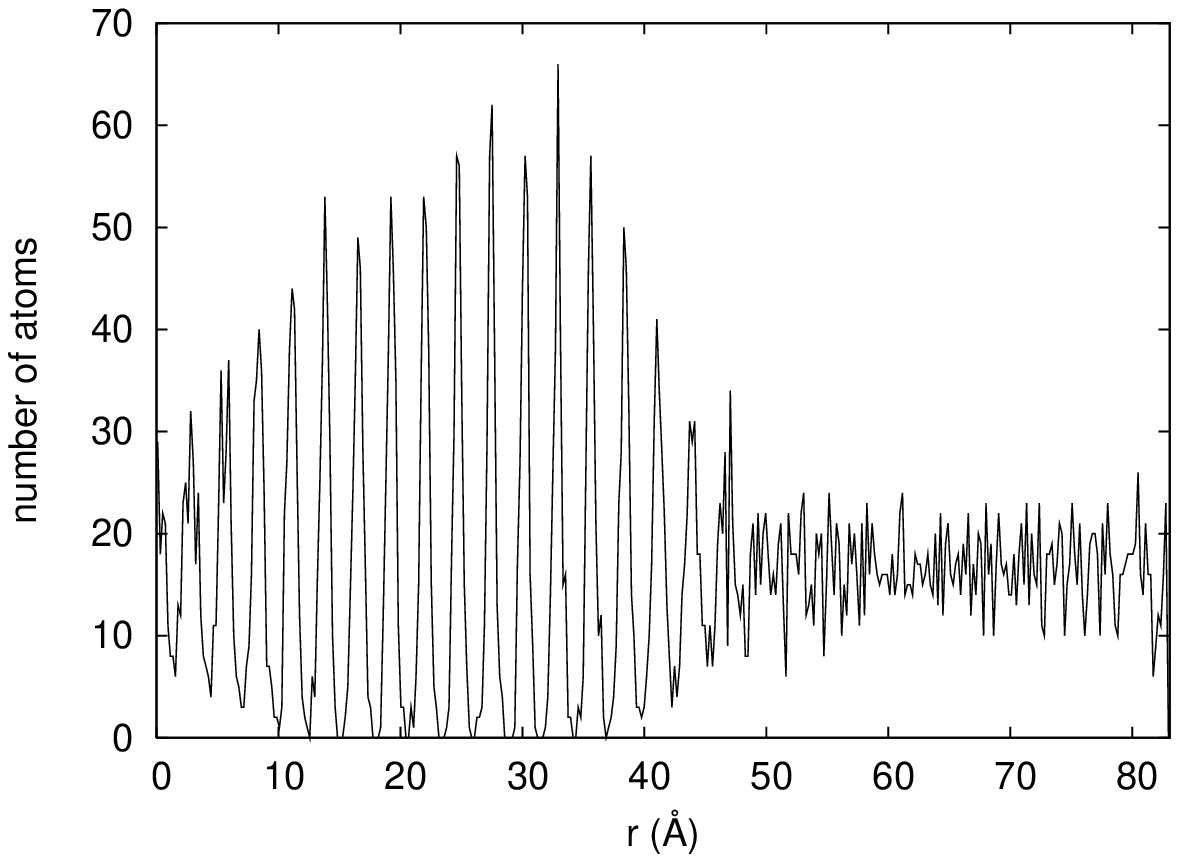}}%
        \caption{}
\label{fig:slice}
\end{figure}

\begin{figure}[c]
\centerline{
        \includegraphics[width=0.8\linewidth]{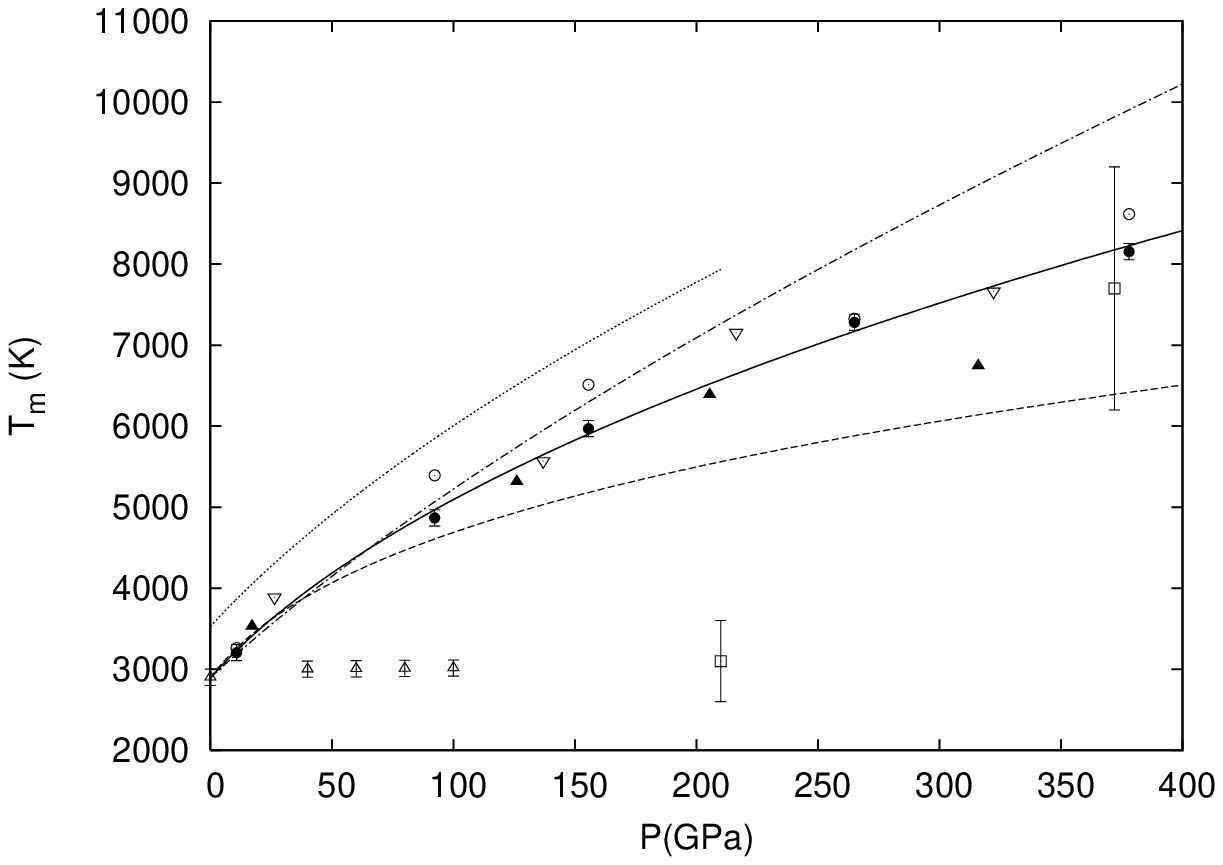}}%
        \caption{}
\label{fig:melting}
\end{figure}

FIG.~\ref{fig:slice}: Density profile in simulation of coexisting solid and liquid Mo at $P = 11$~GPa,
                 $T = 3260$~K after 60~ps. The simulation is performed with the embedded-atom reference
                 model on a system of 6750 atoms.\\

FIG.~\ref{fig:melting}: Calculated \emph{ab initio} melting curve (filled circles and solid line) of this work
                 compared with previous results: generalized pseudopotential calculations of
                 Moriarty~\cite{moriarty94}(dotted line), dislocation-mediated models of
                 Belonoshko \emph{et al.}~\cite{belonoshko04}(long-dashed line) and
                 Verma \emph{et al.}~\cite{verma04}(dashed-dotted line);
                 experimental shock-wave~\cite{hixson89} and DAC~\cite{errandonea01}
                 measurements are shown with empty squares and triangles, respectively.
                 Filled and inverted-empty triangles show solid and liquid
                 \emph{ab initio} molecular dynamics calculations of
                 Belonoshko \emph{et al.}~\cite{belonoshko04}, respectively.
                 Empty circles show results of this work obtained with the
                 EAM model without free-energy corrections.\\

\begin{figure}[c]
\centerline{
        \includegraphics[width=0.8\linewidth]{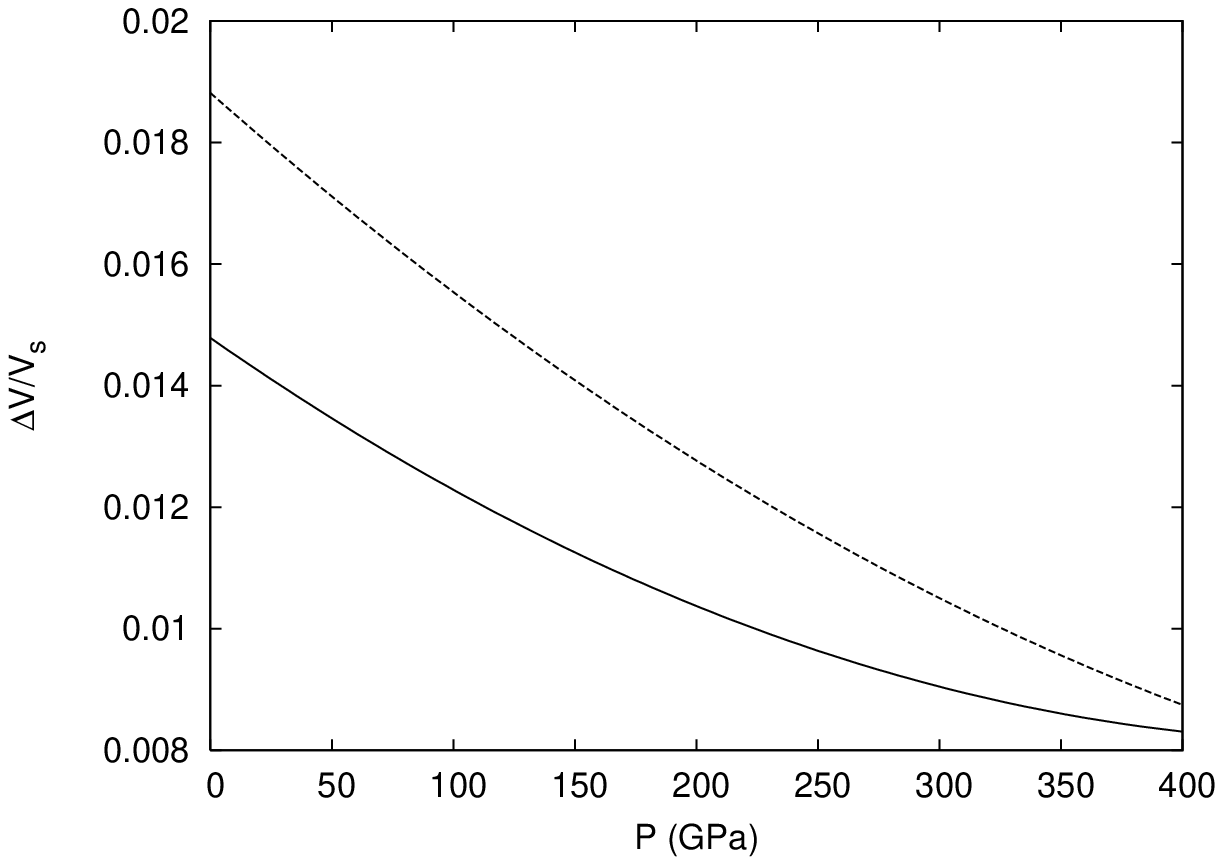}}%
        \caption{}
\label{fig:changevol}
\end{figure}

FIG.~\ref{fig:changevol}: \emph{Ab initio} fractional volume change on melting of Mo as a function of pressure. Solid
                 and dashed curves: present work, with and without free-energy correction, respectively.\\

\begin{figure}[c]
\centerline{
        \includegraphics[width=0.8\linewidth]{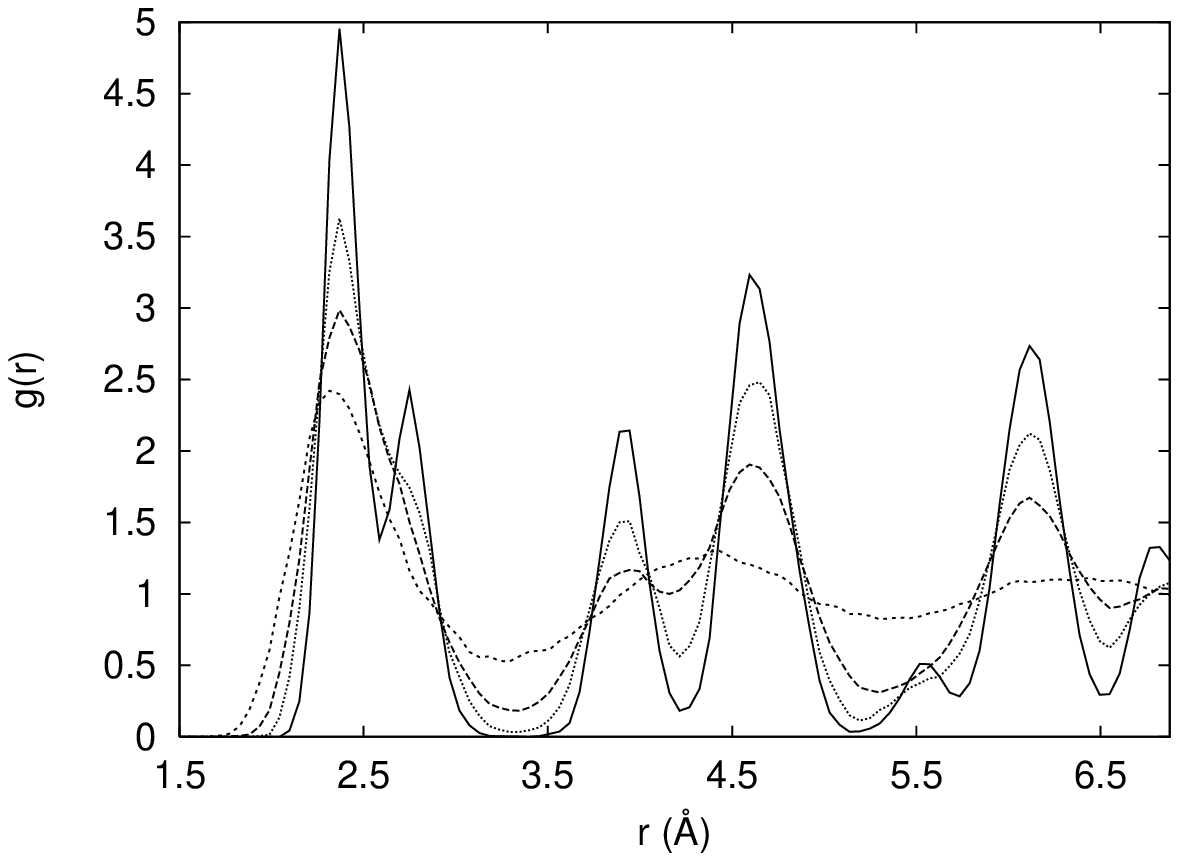}}%
        \caption{}
\label{fig:gr}
\end{figure}

FIG.~\ref{fig:gr}: Calculated radial distribution function of Mo for:
                  solid at $P = 212$~GPa and $T = 2000$~K (solid line),
                  solid at $P = 216$~GPa and $T = 4000$~K (dotted line),
                  solid at $P = 230$~GPa and $T = 6000$~K (long-dashed line)
                  and liquid at $P = 224$~GPa and $T = 7500$~K (short-dashed line).\\

\begin{figure}[c]
\centering
       { \includegraphics[width=0.48\linewidth]{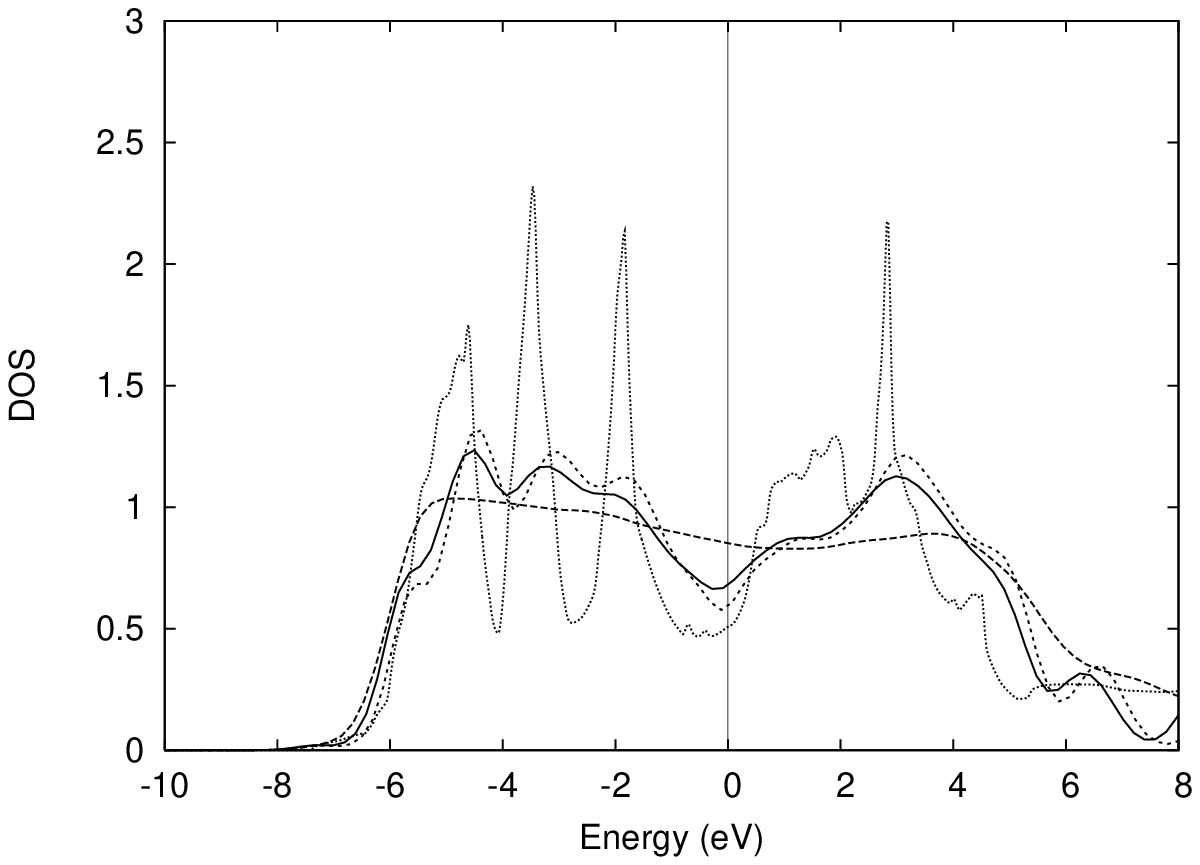} }%
       { \includegraphics[width=0.48\linewidth]{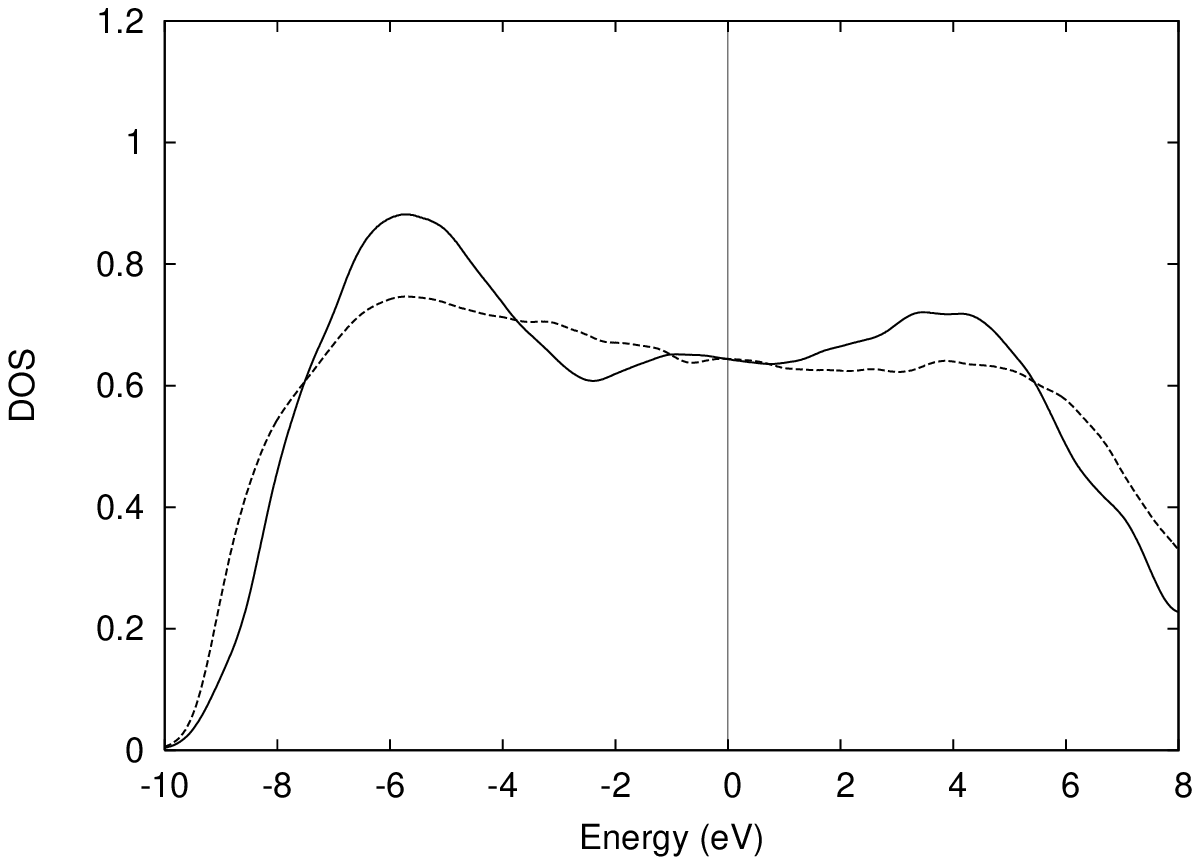} }%
\caption{} 
\label{fig:dosmelt}
\end{figure}

FIG.~\ref{fig:dosmelt}: Density of valence electronic states of Mo at finite temperature and on melting. Left:
                 solid at $P = 48$ GPa and $T = 0$ K (dotted line),
                 solid at $P = 50$ GPa and $T = 2000$ K (short-dashed line),
                 solid at $P = 51$ GPa and $T = 3300$ K (solid line)
                 and liquid at $P = 72$ GPa and $T = 5000$ K (long-dashed line).
                 Right:
                 solid at $P = 285$ GPa and $T = 7000$ K (solid line) and
                 liquid at $P = 300$ GPa and $T = 8250$ K (dashed line).
                 Fermi energy levels are shifted to zero. \\

\clearpage

\clearpage

\clearpage

\clearpage

\clearpage

\clearpage

\clearpage

\clearpage

\clearpage

\clearpage

\clearpage

\end{document}